\documentclass[12pt, a4paper]{article}
\usepackage{amsfonts, amssymb}
\usepackage{graphicx}
\usepackage{latexsym,amsmath,color,array}

\usepackage{natbib}
\usepackage{enumitem}

\usepackage{bbding}
\usepackage{amssymb}
\usepackage{amsmath}
\usepackage{graphicx}
\usepackage{amsmath}
\usepackage{bbm}
\usepackage{mathtools}
 \usepackage{color}
 \usepackage{array}
 \usepackage{multirow}
 \usepackage{makecell}
 \usepackage{colortbl}
 
 \usepackage{subcaption}
  \usepackage{booktabs}
 \usepackage{threeparttable}
 \usepackage{algorithm}
\usepackage{textcomp}
\usepackage{appendix}
\usepackage{xcolor}
\definecolor{mygreen}{rgb}{0.13, 0.55, 0.13} 
\definecolor{mycodebrown}{rgb}{0.64, 0.08, 0.08} 
\definecolor{mycodegreen}{rgb}{0.00, 0.50, 0.00} 
\definecolor{mycodeblue}{rgb}{0.00, 0.00, 1.00} 

\usepackage{relsize} 
\usepackage{listings}
\usepackage{placeins}
\usepackage{setspace}

\lstdefinelanguage{R1}{
    basicstyle=\small \ttfamily,%
    stringstyle=\color{mycodebrown},%
    keywordstyle=\color{mycodeblue},%
    commentstyle=\color{mycodegreen},%
    morekeywords={true, false},
    sensitive=false, 
    showspaces=false,%
    showstringspaces=false,
    basewidth = 0.5em,%
    aboveskip = \medskipamount,%
    belowskip = \medskipamount,%
    morecomment=[l]\#,%
    morestring=[d]",%
    morestring=[d]'
} %

\usepackage{afterpage}

\def\1{\mathbb{I}}

\renewcommand{\thesection}{\arabic{section}}

\newcounter{thm}[section]
\newcounter{appen}[section]
\newcounter{assum}[section]

\begin{document}

\title{Robust Distributional Regression with\\Automatic Variable Selection}
\author{Meadhbh O'Neill\footnote{Department of Mathematics and Statistics, University of Limerick, Limerick, Ireland} \hspace{2cm}
Kevin Burke\footnotemark[1]}
\date{}

\maketitle
\begin{abstract}
Datasets with extreme observations and/or heavy-tailed error distributions are commonly encountered and should be analyzed with careful consideration of these features from a statistical perspective. Small deviations from an assumed model, such as the presence of outliers, can cause classical regression procedures to break down, potentially leading to unreliable inferences. Other distributional features, such as heteroscedasticity, can be handled by going beyond the mean and modelling the scale parameter in terms of covariates. We propose a method that accounts for heavy tails and heteroscedasticity through the use of a generalized normal distribution (GND). The GND contains a kurtosis-characterizing shape parameter that moves the model smoothly between the normal distribution and the heavier-tailed Laplace distribution --- thus covering both classical and robust regression. A key component of statistical modelling is determining the set of covariates that influence the response variable. While correctly accounting for kurtosis and heteroscedasticity is crucial to this endeavour, a procedure for variable selection is still required. For this purpose, we use a novel penalized estimation procedure that avoids the typical computationally demanding grid search for tuning parameters. This is particularly valuable in the distributional regression setting where the location and scale parameters depend on covariates, since the standard approach would have multiple tuning parameters (one for each distributional parameter). We achieve this by using a ``smooth information criterion" that can be optimized directly, where the tuning parameters are fixed at $\log(n)$ in the BIC case.

\smallskip


\end{abstract}

\qquad

\newpage

\section{Introduction} \label{sec:sgnd_introduction}
\begin{figure}[t!]
\centering
\makebox{\includegraphics[width = \textwidth]{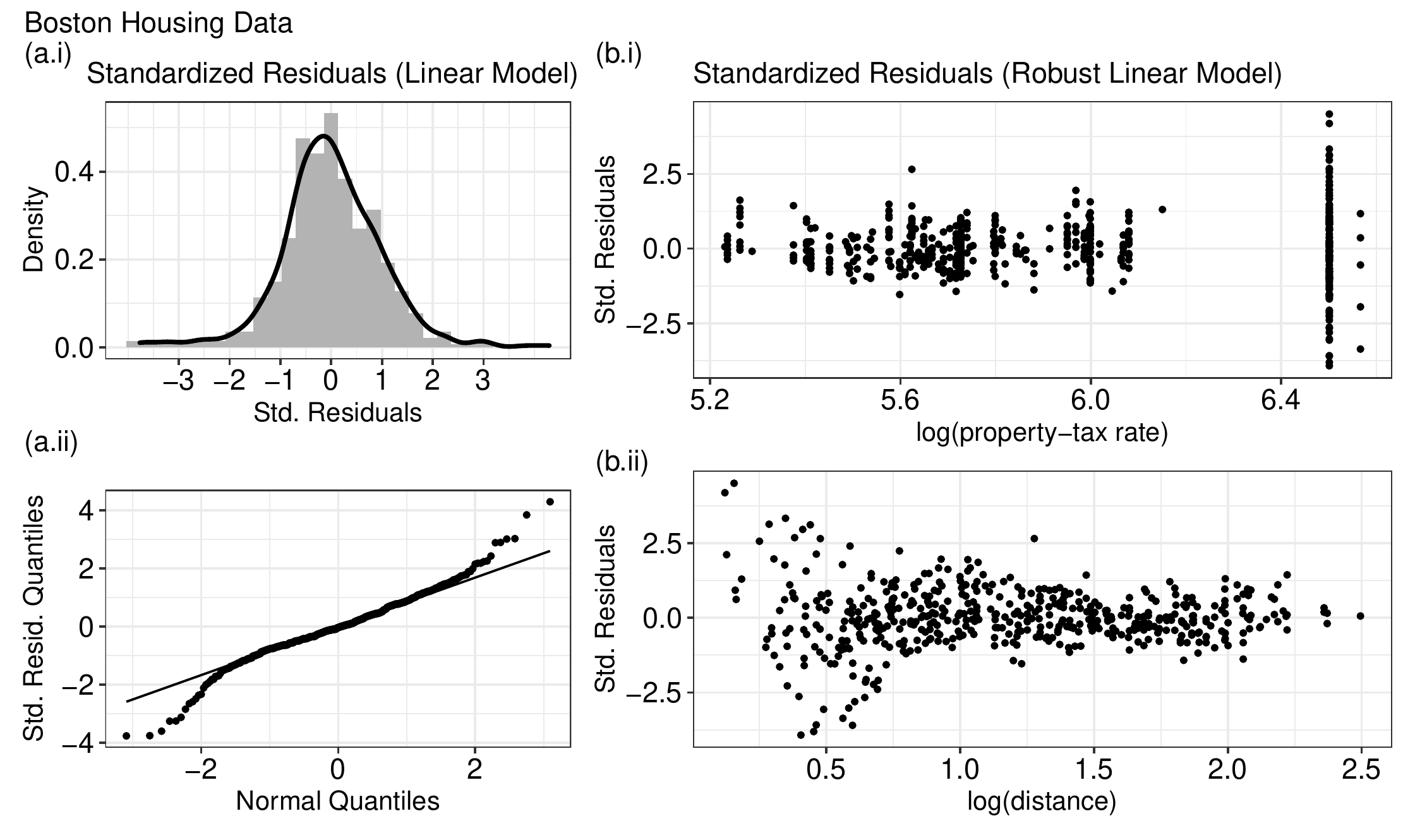}}
\caption{\label{figs:sgnd_hprice}Boston Housing Data: density (a.i) and Q-Q (a.ii) plot of the standardized residuals from fitting a linear model. Standardized residuals from fitting a robust linear model plotted against the independent variables log(property-tax rate) (b.i) and log(distance) (b.ii).}
\end{figure}

Complex datasets with outliers in the response and/or heavy-tailed errors can pose a challenge for standard modelling techniques. Classical regression procedures that assume normally distributed errors minimize the sum of squared errors (SSE) function, which can be prone to pursuing specific observations that differ from the majority of data. These influential observations can have a substantial impact and can distort parameter estimates due to their large residuals \citep{kuhn13applied}. As an example, there is evidence of outliers in the Boston housing data \citep{harrison78hprice2}, which examines the association between median house prices and numerous community characteristics. This is a well known motivating example dataset in the context of normal linear regression, though it appears that the normality assumption on the error distribution may not be appropriate. The density and the quantile-quantile (Q-Q) plot of the standardized residuals from a linear model are shown in Figure~\ref{figs:sgnd_hprice}(a.i) and \ref{figs:sgnd_hprice}(a.ii), which both suggest non-normality of the error distribution. Furthermore, a Shapiro-Wilk test on these residuals leads to a rejection of the null hypothesis of normality (p-value $< 10^{-8}$). Examining the kurtosis of the error distribution implies that the data have a leptokurtic distribution (i.e.,~heavy-tailed relative to a normal distribution). Indeed, a bootstrapped 95\% confidence interval of the excess kurtosis does not contain zero $[1.25,3.75]$.

The presence of small deviations from the model, like this heavy-tailed error distribution, can cause classical regression procedures to become biased and/or inefficient. Therefore, analysis of real data applications using classical methods can be misleading \citep{yu13stability, hanin21cavalier}. When strict model assumptions are violated, the inferential quality disimproves (e.g.,~standard errors and confidence intervals) and can become invalid \citep{ronchetti20accurate}. Robust statistical procedures are extensions of parametric statistics that remain reliable in the presence of these deviations. These methods provide a middle ground between the strictness of a parametric model and the full nonparametric approach, which can be difficult to interpret. Consequently, robust approaches benefit from the simplicity of the parametric structure, both in terms of ease of computation and interpretation, while also having the capability to protect against distributional deviations \citep{avella15robustreview}. A common approach to robust regression is to replace the SSE function in classical regression, which is highly sensitive to outliers, with the sum of the absolute errors function. This is the least absolute deviation (LAD) regression \citep{bassett78asymptotic} and is equivalent to assuming a Laplace error distribution. This is also referred to as $L_1$ regression, and is a type of quantile regression \citep{koenker78quantiles}, where the conditional median of the response variable is estimated. \citet{dielman05least} provides a review of research concerning LAD regression, with \citet{avella15robustreview} delivering a selective overview of different estimators in robust statistics. There are several software implementations available, for example, in \texttt{R} \citep{citer}. The \texttt{rlm} function in the \texttt{MASS} package \citep{massbook} fits a linear model by robust regression that is solved using iteratively reweighted least squares (IRLS). The \texttt{quantreg} package \citep{quantregbook} performs a quantile regression using the \texttt{rq} function and uses linear programming to solve the problem; the \texttt{robustbase} package also uses linear programming via the \texttt{lmrob} function \citep{robustbasepaper, maronna19robustbook}. The \texttt{L1pack} package \citep{L1pack} provides the option to use either IRLS or linear programming to fit a robust linear model using the \texttt{lad} function.

Continuing with our motivating example, we fit a robust linear model to the Boston housing data (via the \texttt{rlm} function in the \texttt{MASS} package). While this will capture the aforementioned heavy tails in the error distribution, we also observe indications of heteroscedasticity in the data. Figure~\ref{figs:sgnd_hprice}(b) displays the standardized residuals from fitting a robust linear model plotted against two of the covariates: log(property-tax rate) and log(distance). These are the log of the full-value property-tax rates and the log of the distance to employment centres respectively. In both cases, it is clear that the variance of the residuals is non-constant with respect to the covariates: the variability increases with the value of log(property-tax rate) and decreases with the value of log(distance). Traditional regression methods tend to focus solely on modelling the mean, which is only one feature of the response distribution. Other potential distributional aspects, such as variability and kurtosis, are often treated as little more than nuisance parameters. This can lead to incomplete analyses, particularly when important dynamics are neglected. In the case of the Boston housing data, the presence of heteroscedasticity indicates that more complex relationships are present, with important insights to be gained by looking beyond the mean. Distributional regression is the approach of simultaneously modelling multiple (or even all) distributional parameters (i.e.,~location, scale and shape) by allowing them to depend on covariates. This reduces the risk of false conclusions and information loss that can be encountered when the conditional distribution is reduced to a single-valued functional (i.e.,~the mean) \citep{kneib13beyond, kneib21rage}. The generalized additive models for location, scale and shape (GAMLSS) model class \citep{rigby05, stas18gamlss} is one of the most popular examples of distributional regression, and is implemented in \texttt{R} using the \texttt{gamlss} package \citep{stas07}. Examination of regression effects beyond the mean, using distributional regression methods such as GAMLSS, presents an opportunity to gain a more comprehensive understanding of the problem at hand.

When there are multiple different covariates that potentially influence the response variable (e.g.,~the Boston housing data has 12), it is useful to perform variable selection to determine a smaller subset of covariates whose effects are sufficiently strong to be deemed important. Classical variable selection techniques, such as stepwise or subset selection, can become computationally expensive; this has, in part, led to the popularity of penalization methods, such as the LASSO (least absolute shrinkage and selection operator) \citep{tibshirani96}, where parameter estimation and model selection is executed simultaneously. The LAD-LASSO \citep{wang07robust} is a robust version of the LASSO, where the quadratic loss function is replaced by the LAD function. There are many extensions of the LAD-LASSO, including methods based on ranks \citep{leng10ladlasso}, adaptive penalized estimation procedures \citep{fan14robustadaptive}, weighted approaches \citep{jiang21outlier} and applications in the high dimensional setting \citep{belloni11, fan17estimation}. \citet{smucler17robust} and \citet{filzmoser21robust} present comprehensive overviews of robust variable selection methods, but current research in this area appears to deal only with the location component of the model, while the variance is assumed constant. To the best of our knowledge, \citet{stas07} were the first to address the variable selection problem with multiple distributional parameters, where they implement stepwise procedures in GAMLSS (see also \citet{stasinopoulos17gamlssbook} and \citet{ramires21gamlssstep}); although GAMLSS is not specific to robust regression, it includes the Laplace distribution. Furthermore, \citet{mayr12} employ gradient boosting techniques to fit GAMLSS models and carry out variable selection; this is implemented in the \texttt{gamboostLSS} package \citep{gamboostLSS}. \citet{hambuckers18understanding} utilize $L_1$-type penalties in the high dimensional distributional regression setting for numeric covariates. \citet{groll19} extend this work by including strategies for nominal and ordinal categorical predictors and present their corresponding package, \texttt{bamlss} \citep{umlauf18bamlsspaper}. However, since there are penalties for multiple distributional parameters, optimal tuning parameter selection (via multidimensional grid search) is computationally intensive, which is a drawback of this otherwise flexible method.

Robust distributional regression with automatic variable selection is an underdeveloped area. We propose a method that covers both classical and robust regression, includes scale regression in addition to location regression, and carries out automatic variable selection. Our starting point is the use of the generalized normal distribution (GND), which contains an additional distributional parameter to be estimated (beyond location and scale) that controls the shape of the distribution and can be regarded as a measure of the kurtosis \citep{box73pe, nelson91conditional, nadarajah05generalized}. Moreover, we adopt a distributional regression approach by allowing both the location and scale to depend on covariates. To carry out automatic variable selection, we extend the work of \citet{oneill2021smoothicarxiv}, who considered normally distributed errors only. Their approach involves the direct optimization of an information criteria, which is particularly advantageous from a computational viewpoint in the distributional regression setting, since it obviates the need for tuning parameter selection. To make our approach amenable to gradient-based optimization (e.g.,~Newton-Raphson), we employ a smooth approximation to both the $L_0$ norm in the information criterion and the absolute value function present in the GND density function. Our proposed method is implemented in the \texttt{smoothic} package \citep{smoothic_package}.

The rest of this paper is organized as follows. Section~\ref{sec:sgnd_methodology} outlines the proposed modelling framework and provides details of the estimation procedure and optimization algorithm. The results from extensive simulation studies are presented in Section~\ref{sec:sgnd_simulation_studies}, followed by the application of the method to two real datasets in Section~\ref{sec:sgnd_real_data_analyses}. Finally, Section~\ref{sec:sgnd_discussion} provides a discussion and some concluding remarks.

\section{Methodology} \label{sec:sgnd_methodology}
\subsection{Preliminaries} \label{sec:sgnd_preliminaries}
The generalized normal distribution (also known as the exponential power distribution or generalized error distribution) generalizes the Laplace and normal distributions through a shape parameter $\kappa$ \citep{box73pe, nelson91conditional, nadarajah05generalized}. The generalized normal distribution (GND) for the response variable $y$ has density function
\begin{equation*}
    f(y) = \frac{\kappa}{2 s \Gamma(1 / \kappa)} e^{-\left(\frac{\lvert y-\mu \rvert}{s}\right)^{\kappa}},
\end{equation*}
where $\mu$, $s>0$, $\kappa>0$ are the location, scale and shape parameters respectively, and $\Gamma(\cdot)$ is the gamma function. This corresponds to the Laplace distribution when $\kappa = 1$ and the normal distribution when $\kappa = 2$. Note that it can be written in the general form
\begin{equation*}
    f(y) = c(\kappa)\frac{1}{s} e^{-g\left(\frac{y-\mu}{s}\right)},
\end{equation*}
where $c(\kappa)=\kappa/2\Gamma(1/\kappa)$ is the normalizing constant and $g(z)=\lvert z \rvert ^{\kappa}$. We suggest using a smooth extension of the absolute value defined as $a_\tau(z) = \sqrt{z^2 + \tau^2} - \tau$, which is differentiable for $\tau > 0$ and $\lim_{\tau\to 0} a_\tau(z) = \lvert z \rvert$. Therefore, replacing $\lvert z \rvert$ with $a_\tau(z)$ in the GND density, yields the smooth GND (SGND) density
\begin{equation}\label{eq:sgnd_density}
    \tilde f(y) = \tilde{c}_{\tau}(\kappa)\frac{1}{s} e^{-\tilde{g}\left(\frac{y-\mu}{s}\right)},
\end{equation}
where $\tilde g(z) = \{a_\tau(z)\}^\kappa$ and $\tilde{c}_{\tau}(\kappa)=1/\int_{-\infty}^{\infty} e^{-\tilde{g}(z)} \,dz$ is a normalizing constant that can be approximated, for example, using numerical integration.

Following the distributional regression framework, the log-likelihood function for the SGND model is
\begin{equation*}
    \ell(\theta)=n\log(\tilde{c}_{\tau}(\kappa)) - \sum_{i=1}^{n} \log(s_i)-\sum_{i=1}^{n}\tilde{g}\left(\frac{y_i-\mu_i}{s_i}\right)
\end{equation*}
with $\mu_i = x_i^T \beta$, $\log(s_i^2)=x_i^T \alpha$ and $\log(\kappa - \kappa_\text{min})=\nu_0$, where the log-link functions are used to ensure the positivity of the parameters $s$ and $\kappa$. The slightly different link function for $\kappa$ is used to restrict $\kappa > \kappa_\text{min}$, for $\kappa_\text{min} > 0$ (e.g.,~$\kappa_\text{min} = 0.2$), to avoid less realistic sub-Laplace distributions with very small $\kappa$ values. (We introduce this mainly for computational reasons, as we have found that optimization can become unstable when $\kappa$ moves into this region.) The response variable is $y_i$ and $x_i=(1, x_{1i}, \ldots, x_{pi})^T$ is the vector of covariates for the $i$th individual; we assume that the covariates are scaled so that they have unit variance, since we will be carrying out penalized estimation. The vector of regression coefficients for the location and scale parameters are ${\beta}=(\beta_{0},\beta_{1},\ldots,\beta_{p})^{T}$ and ${\alpha}=(\alpha_{0},\alpha_{1},\ldots,\alpha_{p})^{T}$. Although $\kappa$ could potentially also depend on covariates, we maintain it as a covariate-independent ``distribution choosing'' parameter here. Moreover, in a somewhat related setting, \citet{burke20} have found that modelling these higher-order distributional shape parameters in terms of covariates may not be required in general (and can sometimes lead to unstable estimation). Thus, the complete parameter vector is $\theta=(\beta^T, \alpha^T, \nu_0)^T$.

In practice, we would generally expect that different sets of covariates will appear in the location and scale components, and this can be dealt with by setting various regression coefficients to zero --- which can be achieved through variable selection. Following the approach of \citet{oneill2021smoothicarxiv}, we achieve automatic variable selection by directly optimizing a ``smooth information criterion" (SIC). This formulates the estimation problem as a penalized likelihood, yielding our proposed SGND-SIC objective function
 \begin{equation}
    \ell^\text{SIC}(\theta)=\ell(\theta)-\frac{\lambda}{2}\left[\lvert \lvert \tilde\beta \rvert \rvert_{0, \epsilon}+ \lvert \lvert \tilde\alpha \rvert \rvert_{0, \epsilon}+3\right],
    \label{eq:sgnd_sic}
\end{equation}
where fixing $\lambda=2$ or $\lambda=\log(n)$ is equivalent to the AIC and BIC (up to a factor of two), respectively, $\lvert \lvert \theta \rvert \rvert_{0, \epsilon} = \sum_{j=1}^{p}\phi_\epsilon(\theta_j)$ is a ``smooth $L_0$ norm", where $\phi_\epsilon(\theta_j) = {\theta_j^2}/({\theta_j^2 + \epsilon^2)}$ and $\epsilon$ is the penalty smoothing parameter such that smaller values of $\epsilon$ bring $\lvert \lvert \theta \rvert \rvert_{0, \epsilon}$ closer to the $L_0$ norm. In this paper, we apply a BIC-type criterion where $\lambda = \log(n)$ throughout. The intercepts are not penalized and are therefore omitted from $\tilde\beta=(\beta_1,\ldots,\beta_{p})^T$ and $\tilde\alpha=(\alpha_1,\ldots,\alpha_{p})^T$. There is an addition of three in the penalty to include the cost of estimating the intercept terms $\beta_0$, $\alpha_0$ and $\nu_0$; this constant is not required for optimization purposes but is included to keep the approach in line with an information criteria. Using a smooth approximation to both the absolute value in the GND and the $L_0$ norm in the information criterion results in a fully differentiable problem. This means that standard gradient based optimization procedures, such as Newton-Raphson, can be used to optimize our smooth objective function as described in the following section.

\subsection{Estimation Procedure}\label{sec:sgnd_estimation_procedure}
We define the penalized estimator as
\begin{equation*}
    \hat{\theta} = {\arg\max}\left(\ell^{\text{SIC}}(\theta)\right),
\end{equation*}
where $\ell^{\text{SIC}}(\theta)$ is given by \eqref{eq:sgnd_sic}. The first derivatives with respect to the parameters are
    \begin{align*} 
            	\begin{split}	
            	\frac{\partial \ell^\text{SIC}}{\partial \beta} &= \frac{\partial \ell}{\partial \beta} - \frac{\log(n)}{2}\xi_{\beta} = X^T z_{\beta} - \frac{\log(n)}{2}\xi_{\beta},\\
            	\frac{\partial \ell^\text{SIC}}{\partial \alpha} &= \frac{\partial \ell}{\partial \alpha} - \frac{\log(n)}{2}\xi_{\alpha} = X^T z_{\alpha} - \frac{\log(n)}{2}\xi_{\alpha},\\
            	\frac{\partial \ell^\text{SIC}}{\partial \nu} &= \frac{\partial \ell}{\partial \nu} = 1_n^T z_{\nu},
            	\end{split}
    \end{align*}
where $X$ is an $n \times (p+1)$ matrix, whose $i$th row is $x_i$; $1_n = (1, \ldots, 1)^T$ is a vector of length $n$; $z_\beta$, $z_\alpha$ and $z_\nu$ are vectors of length $n$ such that
    \begin{align*}
            	\begin{split}	
            	z_{\beta, i} &= \frac{s_{i}^{-2}\kappa(y_i-\mu_{i})a_{i}^{\kappa-1}}{a_{i} + \tau},\\
            	z_{\alpha, i} &= \frac{s_{i}^{-2}\kappa(y_i-\mu_{i})^{2}a_{i}^{\kappa-1}}{2a_{i}+\tau} -\frac{1}{2},\\
            	z_{\nu, i}   &=\frac{\partial}{\partial \nu}\log(\tilde{c}_{\tau}(\kappa))-\log (a_{i})a_{i}^{\kappa}(\kappa - \kappa_\text{min})
            	\end{split}
    \end{align*}
where $a_{i} = \sqrt{s_{i}^{-2}(y_i-\mu_{i})^{2}+\tau^{2}}-\tau$, and $\partial \log(\tilde{c}_{\tau}(\kappa))/\partial \nu$ is given in Appendix~\ref{app:sgnd_derivatives_ctilde}; lastly, the $(j+1)$th elements of the vectors $\xi_\beta$ and $\xi_\alpha$ are given by $\phi^\prime_\epsilon(\beta_j)$ and $\phi^\prime_\epsilon(\alpha_j)$, where $\phi^\prime_\epsilon(\theta_j) = 2\theta_j \epsilon^2/(\theta_j^2+\epsilon^2)^2$, but the first elements are zero because the intercepts are not penalized, i.e., $\phi^\prime_\epsilon(\beta_0) = \phi^\prime_\epsilon(\alpha_0) = 0$. Then $-\nabla_{\theta}\nabla_{\theta}^T\ell^{\text{SIC}}(\theta)$ is given by
\begin{align*}
    I(\theta)
    &= I_0(\theta) +
    \begin{pmatrix}
             \log(n)\Sigma_{\beta}/2
             & 0 & 0 \\
             0 & \log(n)\Sigma_{\alpha}/2 & 0 \\
             0 & 0 & 0
         \end{pmatrix}\\
    &=   \begin{pmatrix}
             X^{T} W_{\beta} X + \log(n)\Sigma_{\beta}/2
             & X^{T} W_{\beta\alpha} X & {1}_n^TW_{\beta \nu}{1}_n\\
             X^{T} W_{\beta\alpha} X & X^{T} W_{\alpha} X + \log(n)\Sigma_{\alpha}/2 & {1}_n^TW_{\alpha \nu}{1}_n \\
             {1}_n^TW_{\beta \nu}{1}_n & {1}_n^TW_{\alpha \nu}{1}_n & {1}_n^TW_{\nu}{1}_n
         \end{pmatrix}
    \end{align*}
where $I_0(\theta) = -\nabla_{\theta}\nabla_{\theta}^T\ell(\theta)$ is the observed information matrix associated with the unpenalized likelihood (i.e.,~\eqref{eq:sgnd_sic} with $\lambda = 0$), $\Sigma_\beta$ and $\Sigma_\alpha$ are diagonal matrices that are present due to the penalties and whose $(j+1)$th diagonal elements are $\phi_\epsilon^{\prime\prime}(\beta_{j})$ and $\phi_\epsilon^{\prime\prime}(\alpha_{j})$, where $\phi_{\epsilon}^{\prime \prime}(\theta_j)=2\epsilon^2(\epsilon^2-3\theta_j^2)/(\theta_j^2+\epsilon^2)^3$, but the first diagonal elements are zero because the intercepts are not penalized, i.e., $\phi_{\epsilon}^{\prime \prime}(\beta_0)=\phi_{\epsilon}^{\prime \prime}(\alpha_0)=0$; $W_\beta$, $W_\alpha$, $W_\nu$, $W_{\beta \alpha}$, $W_{\beta \nu}$ and $W_{\alpha \nu}$  are $n \times n$ diagonal weight matrices whose $i$th diagonal elements are provided in Appendix~\ref{app:sgnd_derivatives_second}.

For the purpose of optimization, we use the ``RS" algorithm \citep{rigby05}, where the off-diagonal elements of the information matrix are set to zero. One justification for this approach is due to parameter orthogonality in location-scale models \citep{cox87}, where the location parameter is information orthogonal to both the scale and shape parameters. However, the scale and shape parameters in this model are not orthogonal \citep{jones11orthog}. Despite this, we have found that the algorithm works well without the cross derivatives; this finding is also supported by \citet{stas07}. Thus, the system of Newton-Raphson equations is
\begin{multline}
    \hspace{-0.4cm}
                \begin{pmatrix}
                     X^{T} W_{\beta}^{(m)} X + \log(n)\Sigma_{\beta}^{(m)}/2
                     & 0 & 0\\
                     0 & X^{T} W_{\alpha}^{(m)} X + \log(n)\Sigma_{\alpha}^{(m)}/2 & 0\\
                     0 & 0 & 1_n^T W_{\nu}^{(m)}1_n
                 \end{pmatrix}
             {
             \begin{pmatrix}
                     \beta^{(m+1)} - \beta^{(m)}\\
                     \alpha^{(m+1)} - \alpha^{(m)}\\
                     \nu^{(m+1)} - \nu^{(m)}
                 \end{pmatrix}}
             \\=
             \begin{pmatrix}
                     X^T z_{\beta}^{(m)} - \log(n)\xi_{\beta}^{(m)}/2\\
                     X^T z_{\alpha}^{(m)} - \log(n)\xi_{\alpha}^{(m)}/2\\
                     1_n^T z_{\nu}^{(m)}\\
             \end{pmatrix},
         \label{eq:sgnd_nr_compact}
    \end{multline}
and this system is iteratively solved for {$\theta^{(m+1)} = ({\beta^{(m+1)}}^T, {\alpha^{(m+1)}}^T, \nu^{(m+1)})^T$}. Note that the elements superscripted by $(m)$ depend on $\theta^{(m)}$, but this dependence is suppressed for notational convenience.

Our \texttt{smoothic} package provides both the above Newton-Raphson procedure and another based on the in-built \texttt{nlm} function in \texttt{R}; we have found the former to be faster and the latter to be more stable (owing to it being a more sophisticated optimizer). We now describe our suggested initialization for the optimization procedure. For the location parameter, we use the classical ordinary least squares estimates, i.e., $\beta^{(0)} = (X^T X)^{-1} X^T Y$, where $Y = (y_1,\ldots,y_n)^T$ is the vector of response values. We then set the intercept of the scale parameter to be $\log(q^2)$, where $q^2 = \sum_{i=1}^n (y_i - x_i^T \beta^{(0)})^2/(n - p)$ is the classical residual variance estimator in normal linear regression. As in \citet{rutemiller68} and \citet{harvey76}, the remaining elements of the $\alpha^{(0)}$ parameter vector are set to zero, which yields $\alpha^{(0)}=(\log(q^2), 0, \ldots, 0)^T$. We initialize the shape parameter at $\nu_0^{(0)}=\log(2-\kappa_\text{min})$, which corresponds to a distribution with normal-like tails. 

Upon convergence, the standard errors of the estimates are directly acquired by estimating the covariance of the penalized estimates for the true non-zero parameters using the sandwich formula,
\begin{equation}
    \hat{\mathrm{cov}}(\hat{\theta})=\{I(\hat{\theta})\}^{-1} I_0(\hat{\theta})\{I(\hat{\theta})\}^{-1},
    \label{eq:sgnd_sandwich}
\end{equation}
which is a general formula used in penalized regression that \citet{fan01, fan02} have shown to be accurate for moderate sample sizes. \citet{oneill2021smoothicarxiv} have also applied this formula successfully with heteroscedastic data and penalized estimation. Obtaining standard errors in a robust regression setting is a known problem \citep{singh98breakdown, salibian02, croux04robust}, with \citet{simpson92se} defining the concept of standard error breakdown. This occurs if the estimated standard error of the estimates are driven to zero or infinity, and, to the best of our knowledge, this remains an issue for penalized estimation with heavy-tailed heteroscedastic data. Indeed, we experience standard error breakdown for small sample sizes in the heavy-tailed heteroscedastic setting when $\kappa = 1$, with the estimated standard errors being driven to zero for small $\tau = 0.05$ (see the Supplementary Material). However, inferential performance does improve for larger sample sizes, suggesting that \eqref{eq:sgnd_sandwich} is working asymptotically as expected. The issue of standard error breakdown is not observed for larger $\kappa$ values that we have tested, implying that the problem is associated with $\kappa$ values near one (and, indeed, also the even heavier-tailed setting where $\kappa<1$ that is not shown here). Moreover, standard error breakdown is not such an issue in the homoscedastic setting (see the Supplementary Material). To improve the situation with respect to standard error breakdown, we suggest setting $\tau = 0.15$, yielding a smoother absolute value function approximation in the density function given in \eqref{eq:sgnd_density}. Although setting $\tau$ even higher than this can eliminate the issue, it also moves us further away from the desired Laplace-like distribution.

\begin{algorithm}[b!]
\caption{Implementation of the SGND-SIC $\epsilon$-telescope Method}
\label{algo:sgnd_tele}
    \begin{enumerate}[label=\textbf{\arabic*}.]
        \item \textbf{Initialization:} Rescale covariates to have unit variance and set $\theta^{(0)}=(\beta^{(0)T},\alpha^{(0)T}, \nu_0^{(0)})^T$, where $\beta^{(0)}$, $\alpha^{(0)}$ and $\nu_0^{(0)}$ are the initial values for the location, scale and shape parameters respectively (see Section~\ref{sec:sgnd_estimation_procedure}).
        \item \textbf{$\mathbf{\epsilon}$-telescoping:} Run through the sequence of telescope values of length $T$ from $\epsilon_1$ to $\epsilon_T$, where $\epsilon_t=\epsilon_1 r^{t-1}$ for step $t=1,\ldots,T$ and $r \in (0,1)$ is the rate of exponential decay. We suggest $\epsilon_1=10$, $\epsilon_T=10^{-4}$ and $T=100$. 
        \begin{itemize}[leftmargin=*]
        \item\textbf{For $\boldsymbol{t=1,\ldots,T}$:}\\
        \textbf{Optimization:} Maximize $\ell^{\text{SIC}}(\theta)$ in \eqref{eq:sgnd_sic} to obtain $\hat{\theta}_{\epsilon_t}$ by iteratively re-solving the system of equations in \eqref{eq:sgnd_nr_compact} using initial values $\theta^{(0)}_{\epsilon_t}$ (where $\theta^{(0)}_{\epsilon_1}=\theta^{(0)}$). When $\lvert \theta^{(m+1)}_{\epsilon_t} - \theta^{(m)}_{\epsilon_t}\rvert\leq\omega$ for some small tolerance, e.g., $\omega = 10^{-8}$, convergence is achieved. The obtained estimates are used as initial values for the next step in the telescope, i.e., set $\theta^{(0)}_{\epsilon_{t+1}}$ = $\hat{\theta}_{\epsilon_t}$ to exploit warm starts.
        \end{itemize}
        \item \textbf{Output:} The final estimates $\hat{\theta}_{\epsilon_T}$ are acquired at $t=T$ and any estimates that are sufficiently close to zero (e.g.,~below $10^{-5}$) can be treated as being zero. The associated standard errors are calculated by evaluating \eqref{eq:sgnd_sandwich} at $\hat{\theta}_{\epsilon_T}$. The final estimates are converted back to their original scale (by dividing by the covariate sample standard deviations) so that they correspond to the original, unscaled covariates.
    \end{enumerate}
\end{algorithm}

\subsection{Algorithm}\label{sec:sgnd_algorithm}
To induce sparsity, we aim for a small value of $\epsilon$ in the $L_0$ approximation. However, starting immediately with a small $\epsilon$ is challenging from an optimization perspective (since it is very close to the non-differentiable $L_0$ problem). Thus, we suggest an $\epsilon$-telescoping procedure \citep{oneill2021smoothicarxiv} where a sequence of successively smaller $\epsilon$ values are used as a way of providing good starting values (i.e.,~warm starts) towards a final small $\epsilon$. For the purpose of penalized estimation, we rescale covariates to have unit variance, i.e., we use $x_j / SD(x_j)$ in place of $x_j$ where $SD(x_j)$ is the sample standard deviation of $x_j$. Typically, for the final solution, unscaled regression coefficients are preferred. Therefore, we convert the coefficients back to their original scale. Considering this, our overall optimization procedure is summarized in Algorithm~\ref{algo:sgnd_tele}.

\section{Simulation Studies}\label{sec:sgnd_simulation_studies}
\begin{table}[b!]
\caption{\label{tab:sgnd_true_values}True parameter values}%
\centering
\begin{tabular}{@{}cccccccccccccc@{}}
  \toprule
    & & \textcolor{blue}{E} & \textcolor{red}{M} & \textcolor{mygreen}{B} & N & N & \textcolor{red}{M} & N & N & \textcolor{red}{M} & \textcolor{mygreen}{B} & \textcolor{blue}{E} & \textcolor{red}{M} \\
    & $X_0$ & \textcolor{blue}{$X_1$} & \textcolor{red}{$X_2$} & \textcolor{mygreen}{$X_3$} & ${X_4}$ & ${X_5}$ & \textcolor{red}{$X_6$} & ${X_7}$ & $X_8$ & \textcolor{red}{${X_9}$} & \textcolor{mygreen}{$X_{10}$} & \textcolor{blue}{$X_{11}$} & \textcolor{red}{$X_{12}$} \\
 \midrule
$\beta$ & 0 & 1 & 0.5 & 0.5 & 1 & 0.5 & 1 & 0 & 0 & 0 & 0 & 0 & 0 \\[0.05cm] 
$\alpha$ & 0 & 0.5 & 1 & 0.5 & 1 & 0 & 0 & 0.5 & 1 & 0 & 0 & 0 & 0 \\
\bottomrule
\multicolumn{14}{p{0.55\textwidth}}{\footnotesize \textcolor{blue}{E = Exponential}, \textcolor{mygreen}{B = Bernoulli}, N = independent normal, \footnotesize \textcolor{red}{M = multivariate normal (correlated)}.}
\end{tabular}
\vspace{-1cm}
\end{table}

\subsection{Setup}
The performance of our proposed SGND-SIC method is investigated through simulation studies. Data are simulated from the SGND defined in \eqref{eq:sgnd_density}, with $\tau=0.15$ and $\kappa_\text{min} = 0.2$. The regression coefficients are shown in Table~\ref{tab:sgnd_true_values}, where $X$ $(X_1, \ldots, X_{12})$ is the matrix of 12 covariates. The setup of the regression coefficients enable the method to be assessed where covariates enter through both distributional parameters simultaneously, where both the $\beta$ and $\alpha$ coefficients are non-zero ($X_1$ to $X_4$), or through a single distributional parameter, where only the $\beta$ coefficient is non-zero ($X_5$ and $X_6$) or only the $\alpha$ coefficient is non-zero ($X_7$ and $X_8$). Lastly, the pure noise covariates ($X_9$ to $X_{12}$) represent the case where the covariates do not enter the model at all, i.e., both the $\beta$ and $\alpha$ coefficients are zero so that these covariates have no effect. The true values of 0.5 and 1 correspond to covariates having a weaker and stronger effect. The covariates are distributed as follows: $(X_1, X_{11}) \sim$ Exponential(1) corresponding to two skewed covariates; $(X_3, X_{10}) \sim$ Bernoulli(0.75) giving two unbalanced binary covariates; four independent normal covariates $(X_4, X_5, X_7, X_8) \sim$ $\text{N}(0,1)$; and four correlated multivariate normal covariates $(X_2, X_6, X_9, X_{12}) = (Z_1, Z_2, Z_3, Z_4) \sim$ MVN wherein $\text{corr}(Z_j, Z_k)= 0.5^{\lvert j - k \rvert}$. The shape parameter $\kappa$ is varied such that $\kappa \in (1, 1.33, 1.67, 2)$ (rounded to two decimal places), which corresponds to a Laplace-like distribution when $\kappa=1$, a normal-like distribution when $\kappa = 2$, and a distribution between these when $\kappa \in (1.33, 1.67)$. These values of $\kappa$ equate to true values of $\nu_0 \in (-0.22, 0.13, 0.38, 0.59)$ (rounded to two decimal places), where $\nu_0=\log(\kappa-\kappa_\text{min})$. Sample sizes of 500, 1000 and 5000 are considered, where each scenario is replicated 1000 times.

\subsection{Simulation Results}
We assess the variable selection performance in terms of the number of true zero coefficients correctly set to zero (C) and the probability of choosing the true model (PT). We also consider the mean squared error $\text{MSE}(\hat{\theta})=(\hat{\theta} - \theta)^T{X^TX}(\hat{\theta} - \theta)/n$, which is a measure of in-sample prediction accuracy \citep{tibshirani97}. Table~\ref{tab:sgnd_var_selection} displays these metrics averaged over simulation replicates.

\begin{table}[b!]
\caption{Simulation results: model selection metrics} 
\label{tab:sgnd_var_selection}
\centering
\resizebox{\textwidth}{!}{
\begin{tabular}{@{}c@{~~~}l@{~~~}   c@{~~}c@{~~}c@{~~}   c@{~~}   c@{~~}c@{~~}c@{~~}   c@{~~}   c@{~~}c@{~~}c@{~~}   c@{~~}   c@{~~}c@{~~}c@{}}
  \toprule
 {} & {} & \multicolumn{3}{c}{$\kappa = 1$} && \multicolumn{3}{c}{$\kappa = 1.33$} && \multicolumn{3}{c}{$\kappa = 1.67$} && \multicolumn{3}{c}{$\kappa = 2$} \\
 \cmidrule(r){3-5} \cmidrule(r){7-9} \cmidrule(r){11-13} \cmidrule(){15-17}
 {} & $n$ & C(6) & PT & MSE && C(6) & PT & MSE && C(6) & PT & MSE && C(6) & PT & MSE\\ 
 \midrule
 {$\beta$}                & 500   & 5.83 & 0.84 & 0.01 && 5.88 & 0.89 & 0.01 && 5.89 & 0.90 & 0.01 && 5.90 & 0.90 & 0.01 \\ 
 {}                       & 1000  & 5.89 & 0.89 & 0.01 && 5.92 & 0.93 & 0.00 && 5.94 & 0.94 & 0.00 && 5.94 & 0.94 & 0.00 \\ 
 {}                       & 5000  & 5.97 & 0.97 & 0.00 && 5.97 & 0.97 & 0.00 && 5.97 & 0.97 & 0.00 && 5.97 & 0.97 & 0.00 \\[0.2cm]
 
 {$\alpha$}                & 500   & 5.90 & 0.89 & 0.12 && 5.89 & 0.90 & 0.07 && 5.89 & 0.90 & 0.05 && 5.89 & 0.90 & 0.04\\ 
 {}                        & 1000  & 5.94 & 0.94 & 0.05 && 5.94 & 0.94 & 0.03 && 5.94 & 0.94 & 0.02 && 5.94 & 0.94 & 0.02\\ 
 {}                        & 5000  & 5.99 & 0.99 & 0.01 && 5.99 & 0.99 & 0.01 && 5.99 & 0.99 & 0.00 && 5.99 & 0.99 & 0.00\\
   \bottomrule
   \multicolumn{17}{p{0.95\textwidth}}{\footnotesize C, average correct zeros; PT, the probability of choosing the true model; MSE, the average mean squared error.}\\
\end{tabular}}
\end{table}

For all values of $\kappa$, the C values are close to the true value of six for both the location and scale parameters, and this improves as the sample size increases. We also note that the method never incorrectly sets a variable to zero in the scenarios we have considered (although this is not shown in Table~\ref{tab:sgnd_var_selection}). In terms of the probability of choosing the correct model (PT) for both $\beta$ and $\alpha$, we see that this is generally above 90\% (approximately) and increases with the sample size; the only exception is the $\beta$ component when $\kappa = 1$ at the smallest sample size of $n=500$, where PT$=84\%$. The MSE values are generally small and decrease with the sample size.

\begin{table}[t!]
\caption{Simulation results: estimation and inference metrics}
\label{tab:sgnd_parameter_inference_short}
\resizebox{\textwidth}{!}{
\begin{tabular}{@{}l@{~~} c@{~~}c@{~~}  c@{~~}c@{~~}c@{~~}c@{~~}  c@{~~}  c@{~~}c@{~~}c@{~~}c@{~~}  c@{~~}  c@{~~}c@{~~}c@{~~}c@{}}
\toprule
{} & {} & {} & \multicolumn{4}{c}{$n = 500$} && \multicolumn{4}{c}{$n = 1000$} && \multicolumn{4}{c}{$n = 5000$} \\
\cmidrule(r){4-7} \cmidrule(r){9-12} \cmidrule(){14-17}
{$\kappa$} & {} & $\theta$ & $\hat{\theta}$ & SE & SEE & CP && $\hat{\theta}$ & SE & SEE & CP && $\hat{\theta}$ & SE & SEE & CP \\
  \midrule
  1       & $\beta_{1}$   & 1.00  & 1.00  & 0.05 & 0.04 & 0.74 && 1.00  & 0.04 & 0.03 & 0.83 && 1.00  & 0.01 & 0.01 & 0.92 \\
  {}      & $\alpha_{1}$  & 0.50  & 0.51  & 0.10 & 0.09 & 0.94 && 0.50  & 0.07 & 0.06 & 0.94 && 0.50  & 0.03 & 0.03 & 0.94 \\
  {}      & $\nu_{0}$     & -0.22 & -0.22 & 0.11 & 0.10 & 0.90 && -0.22 & 0.07 & 0.07 & 0.92 && -0.22 & 0.03 & 0.03 & 0.92 \\[0.2cm]
  1.33    & $\beta_{1}$   & 1.00 & 1.00  & 0.04 & 0.04 & 0.91  && 1.00  & 0.03 & 0.03 & 0.94  && 1.00  & 0.01 & 0.01 & 0.94 \\
  {}      & $\alpha_{1}$  & 0.50 & 0.51  & 0.08 & 0.08 & 0.94  && 0.50  & 0.06 & 0.05 & 0.94  && 0.50  & 0.02 & 0.02 & 0.94 \\
  {}      & $\nu_{0}$     & 0.13 & 0.16  & 0.11 & 0.10 & 0.92  && 0.14  & 0.07 & 0.07 & 0.92  && 0.13  & 0.03 & 0.03 & 0.93 \\[0.2cm]
  1.67    & $\beta_{1}$   & 1.00 & 1.00  & 0.04 & 0.03 & 0.92 && 1.00  & 0.03 & 0.02 & 0.95 && 1.00  & 0.01 & 0.01 & 0.94 \\
  {}      & $\alpha_{1}$  & 0.50 & 0.50  & 0.07 & 0.07 & 0.94 && 0.50  & 0.05 & 0.05 & 0.95 && 0.50  & 0.02 & 0.02 & 0.94 \\
  {}      & $\nu_{0}$     & 0.38 & 0.43  & 0.12 & 0.11 & 0.92 && 0.40  & 0.08 & 0.07 & 0.94 && 0.38  & 0.03 & 0.03 & 0.95 \\[0.2cm]
  2       & $\beta_{1}$   & 1.00 & 1.00  & 0.03 & 0.03 & 0.92 && 1.00  & 0.02 & 0.02 & 0.95 && 1.00  & 0.01 & 0.01 & 0.95 \\
  {}      & $\alpha_{1}$  & 0.50 & 0.50  & 0.07 & 0.06 & 0.94 && 0.50  & 0.05 & 0.04 & 0.94 && 0.50  & 0.02 & 0.02 & 0.94 \\
  {}      & $\nu_{0}$     & 0.59 & 0.65  & 0.13 & 0.12 & 0.91 && 0.62  & 0.08 & 0.08 & 0.93 && 0.59  & 0.03 & 0.03 & 0.95 \\
  \bottomrule
  \multicolumn{17}{p{0.95\textwidth}}{\footnotesize Full results for all coefficients available in the Supplementary Material; SE, standard deviation of estimates over 1000 replications; SEE, average of estimated standard errors over 1000 replications; CP, the empirical coverage probability of a nominal 95\% confidence interval.}\\
  \end{tabular}}
\end{table}

The estimation and inferential performance of our method is examined in Table~\ref{tab:sgnd_parameter_inference_short}, which only shows results for $\beta_1$, $\alpha_1$ and $\nu_0$ in the interest of brevity (see the Supplementary Material for full results). More specifically, we display the average estimate over simulation replicates, the true standard error (SE) (i.e.,~the standard deviation of the estimates over the replicates), the average estimated standard error (SEE) (computed using \eqref{eq:sgnd_sandwich} in a given replicate), and the empirical coverage probability (CP) for a nominal 95\% confidence interval. In all cases, we can see that the parameter estimates are quite unbiased, apart from some slight bias in $\hat \nu_0$ for the smallest sample size. As for uncertainty quantification, we find that this is generally good: the SEs are well estimated by the SEEs and the confidence interval coverage is near 95\%. The exception to this is $\kappa = 1$ (especially at $n=500$) where the SEs are underestimated and the coverage is poor. However, the results are much improved at $n=1000$ and $n=5000$ suggesting that \eqref{eq:sgnd_sandwich} is asymptotically valid. In any case, the poorer results for $\kappa = 1$ are in line with the discussion in Section~\ref{sec:sgnd_estimation_procedure} about standard error breakdown in the context of robust regression. Of course, an alternative approach would be to estimate the standard errors in a nonparametric manner through bootstrapping; we have found that this arguably improves the situation with conservative SEE values (see the Supplementary Material).

\section{Real Data Analyses} \label{sec:sgnd_real_data_analyses}
\subsection{Overview}\label{sec:sgnd_real_data_analyses_overview}
In this section, we consider our proposed SGND-SIC estimation procedure applied to two datasets, namely, the Boston housing data (Section~\ref{sec:sgnd_real_data_analyses_hprice}) and a diabetes dataset (Section~\ref{sec:sgnd_real_data_analyses_diabetes}), both of which are accessible in the \texttt{smoothic} package \citep{smoothic_package}. More specifically, we model the location and scale parameters in terms of covariates (as described in Section~\ref{sec:sgnd_methodology}), hereafter referring to this as SGND-MPR to highlight the multi-parameter regression nature of the model. We compare this to the SGND-SPR approach, which is a so-called ``single-parameter regression" model wherein only the location parameter depends on covariates, i.e., the variance is a constant; this SGND-SPR model is analogous to standard (normal and robust) regression approaches, albeit with parameters estimated using our novel SGND-SIC procedure. Both models can be easily implemented (by using \texttt{model = "mpr"} or \texttt{"spr"}) with the following code.
\begin{lstlisting}[language=R1]
# SGND-SIC (MPR & SPR) --------------------
library(smoothic)
fit <- smoothic(formula = lcmedv ~ ., # lcmedv is the response variable
                data = bostonhouseprice2,
                family = "sgnd", # Smooth Generalized Normal Distribution
                model = "mpr")   # or model = "spr" for location only
\end{lstlisting}

We also compare our proposal to a variety of other methods, the first of which is the GAMLSS stepwise procedure via the \texttt{gamlss} package \citep{stas07}, and which we refer to as ``GAMLSS-STEP". More specifically, this approach is implemented in the \texttt{stepGAICAll.A()} function, whose argument \texttt{k} we set to $\log(n)$ corresponding to BIC as shown below.
\begin{lstlisting}[language=R1]
# GAMLSS-STEP -----------------------------
library(gamlss)
library(gamlss.dist) # for Power Exponential Distribution (PE2)
# Initial model in the stepwise search
initial_fit <- gamlss(lcmedv ~ 1,
                      data = bostonhouseprice2,
                      family = PE2)
# Strategy A stepwise-based approach
f_gamlssstep <- as.formula("~ crim + zn + indus + rm + age + rad + 
                           ptratio + lnox + ldis + ltax + llstat + chast")
fit <- stepGAICAll.A(initial_fit,
                     scope = list(lower = ~1, # model with no predictors
                                  upper = f_gamlssstep), # all predictors
                     k = log(nrow(bostonhouseprice2)), # BIC k = log(n)
                     nu.try = FALSE) # no covariates in the nu shape parameter
\end{lstlisting}
Note that \texttt{family = PE2} relates to the Power Exponential (PE) distribution (accessed via the \texttt{gamlss.dist} package \citep{rigby05}), where \texttt{PE2} corresponds to the standard parameterization of the distribution. The PE distribution is equivalent to the GND, making it comparable to our proposed SGND-MPR method.

The \texttt{gamboostLSS} package \citep{gamboostLSS} fits a GAMLSS model using distributional component-wise gradient boosting, wherein variable selection is a consequence of stopping the algorithm early. This is controlled by setting the number of iterations using the \texttt{mstop} argument, which is a vector, since the stopping point can be different for each distributional parameter. Choosing an optimal value, say, \texttt{opt\char`_mstop}, is often based on cross-validation, but we use BIC here to be more in line with our proposed BIC-based optimization procedure. (Note that we have found the results from cross-validation minimization to be broadly similar to the results from the BIC-based procedure that are presented in this section.) Due to the algorithmic nature of gradient boosting, there is no commonly accepted method to evaluate the degrees of freedom of a boosting fit \citep{gamboostLSS}; therefore, we take the degrees of freedom as the number of non-zero coefficients. We refer to this as the ``GAMLSS-BOOST" method, and it can be implemented for a specific \texttt{mstop} using the code below (which assumes that an optimal value \texttt{opt\char`_mstop} has already been determined by computing the BIC for a variety of \texttt{mstop} values).
\begin{lstlisting}[language=R1]
# GAMLSS-BOOST ----------------------------
library(gamboostLSS)
library(gamlss.dist) # for Power Exponential Distribution (PE2)
# Assume the optimal stopping iteration (opt_mstop) was found using BIC search
fit <- glmboostLSS(formula = list(mu = lcmedv ~ ., # location
                                  sigma = lcmedv ~ ., # scale
                                  nu = lcmedv ~ 1), # no covariates in shape
                  data = bostonhouseprice2,
                  families = as.families("PE2"),
                  control = boost_control(mstop = opt_mstop)) # early stopping
\end{lstlisting}

The final distributional regression method that we compare to ours is BAMLSS via the \texttt{bamlss} package \citep{umlauf18bamlsspaper}, wherein separate LASSO penalties are applied to each distributional parameter. Note that \texttt{bamlss} selects the LASSO tuning parameters by minimizing BIC using a 100-point grid search for each of the distributional parameters, which is computationally intensive since there are multiple distributional parameters; this is pronounced further by the fact that a penalty appears for the $\nu$ shape parameter even though it is covariate independent, i.e., a $100 \times 100 \times 100$ point grid search is used. The BAMLSS estimates are obtained by running the following code. 
\begin{lstlisting}[language=R1]
# BAMLSS ----------------------------------
library(bamlss)
library(gamlss.dist) # for Power Exponential Distribution (PE2)
f_bamlss <- list(lcmedv ~ la(crim + zn + indus + rm + age + rad + ptratio
                             + lnox + ldis + ltax + llstat + chast),
                 sigma ~ la(crim + zn + indus + rm + age + rad + ptratio
                            + lnox + ldis + ltax + llstat + chast),
                 nu ~ 1)
fit <- bamlss(f_bamlss,
              data = bostonhouseprice2,
              family = PE2,
              optimizer = opt_lasso, # LASSO optimizer using BIC grid search
              criterion = "BIC",
              multiple = TRUE) # separate tuning parameters for each
                               # distributional parameter
\end{lstlisting}

As further comparator models, we also implement the adaptive LASSO (ALASSO) and LAD-LASSO methods, which are both location regression models with constant variance that are commonly used for normal and heavy-tailed scenarios respectively. In line with the other methods, we use the BIC to select penalty tuning parameters. For further details and code, see Appendix~\ref{app:sgnd_additional_code}. 

\subsection{Boston Housing Data}\label{sec:sgnd_real_data_analyses_hprice}
\citet{harrison78hprice2} collected data from 506 suburb areas of Boston, consisting of housing prices and other characteristics. Due to the presence of outliers, analysis of this dataset is prominent in research relating to robust regression \citep{wu09variable, leng10ladlasso, amin15scad, amato21penalised}. The data are available in the \texttt{mlbench} package \citep{mlbench_pkg}. The response variable is the log(corrected median value of owner-occupied homes in \$1000's). There are 12 independent variables: log(full-value property-tax rate per \$10,000) (\texttt{ltax}), average number of rooms per dwelling (\texttt{rm}), log(weighted distances to five Boston employment centres) (\texttt{ldis}), log(percentage of ``lower status" in the population) (\texttt{llstat}), pupil-teacher ratio by town (\texttt{ptratio}), per capita crime rate by town (\texttt{crim}), index of accessibility to radial highways (\texttt{rad}), log(nitric oxides concentration in parts per ten million) (\texttt{lnox}), proportion of owner-occupied units built prior to 1940 (\texttt{age}), Charles River dummy variable (=1 if tract bounds river; 0 otherwise) (\texttt{chast}), proportion of residential land zoned for lots over 25,000 sq.\ ft (\texttt{zn}) and the proportion of non-retail business acres per town (\texttt{indus}).

\begin{table}[t!]
\caption{\label{tab:sgnd_hprice_vars}Boston Housing Data: variables selected and summary metrics}
\centering
\resizebox{\textwidth}{!}{
\setlength{\tabcolsep}{4pt}{
\begin{tabular}{@{}l@{~~} c@{~} c@{~~}c@{~~}c@{~~}c@{~~} c@{~~~} c@{~~}c@{~~}c@{~~}c@{}}
\toprule
{} && \multicolumn{4}{c}{Location \& Scale} && \multicolumn{3}{c}{Location Only} & {}\\
\cmidrule(r){2-6} \cmidrule(l){7-10}
{}  &&   {\begin{tabular}{@{}c@{}}SGND \\ -MPR\end{tabular}} & {\begin{tabular}{@{}c@{}}GAMLSS \\ -STEP\end{tabular}} & {\begin{tabular}{@{}c@{}}GAMLSS \\ -BOOST\end{tabular}} & BAMLSS && {\begin{tabular}{@{}c@{}}SGND \\ -SPR\end{tabular}} & ALASSO & {\begin{tabular}{@{}c@{}}LAD \\ -LASSO\end{tabular}} & Total \\
\midrule
\texttt{ltax}    && \cellcolor{mygreen} \textcolor{white}{$\beta$} & \cellcolor{mygreen} \textcolor{white}{$\beta$} & \cellcolor{mygreen} \textcolor{white}{$\beta$} & \cellcolor{mygreen} \textcolor{white}{$\beta$} && \cellcolor{mygreen} \textcolor{white}{$\beta$} & \cellcolor{mygreen} \textcolor{white}{$\beta$} & \cellcolor{mygreen} \textcolor{white}{$\beta$}  & 7 \\  
\texttt{rm}      && \cellcolor{mygreen} \textcolor{white}{$\beta$} & \cellcolor{mygreen} \textcolor{white}{$\beta$} & \cellcolor{mygreen} \textcolor{white}{$\beta$} & \cellcolor{mygreen} \textcolor{white}{$\beta$} && \cellcolor{mygreen} \textcolor{white}{$\beta$} & \cellcolor{mygreen} \textcolor{white}{$\beta$} & \cellcolor{mygreen} \textcolor{white}{$\beta$}  & 7 \\  
\texttt{ldis}    && \cellcolor{mygreen} \textcolor{white}{$\beta$} & \cellcolor{mygreen} \textcolor{white}{$\beta$} & {} & \cellcolor{mygreen} \textcolor{white}{$\beta$} && \cellcolor{mygreen} \textcolor{white}{$\beta$} & \cellcolor{mygreen} \textcolor{white}{$\beta$} & \cellcolor{mygreen} \textcolor{white}{$\beta$}                                              & 6 \\  
\texttt{llstat}  && \cellcolor{mygreen} \textcolor{white}{$\beta$} & \cellcolor{mygreen} \textcolor{white}{$\beta$} & \cellcolor{mygreen} \textcolor{white}{$\beta$} & \cellcolor{mygreen} \textcolor{white}{$\beta$} && \cellcolor{mygreen} \textcolor{white}{$\beta$} & \cellcolor{mygreen} \textcolor{white}{$\beta$} & \cellcolor{mygreen} \textcolor{white}{$\beta$}  & 7 \\  
\texttt{ptratio} && \cellcolor{mygreen} \textcolor{white}{$\beta$} & \cellcolor{mygreen} \textcolor{white}{$\beta$} & \cellcolor{mygreen} \textcolor{white}{$\beta$} & \cellcolor{mygreen} \textcolor{white}{$\beta$} && \cellcolor{mygreen} \textcolor{white}{$\beta$} & \cellcolor{mygreen} \textcolor{white}{$\beta$} & \cellcolor{mygreen} \textcolor{white}{$\beta$}  & 7 \\  
\texttt{crim}    && \cellcolor{mygreen} \textcolor{white}{$\beta$} & \cellcolor{mygreen} \textcolor{white}{$\beta$} & {} & \cellcolor{mygreen} \textcolor{white}{$\beta$} && \cellcolor{mygreen} \textcolor{white}{$\beta$} & \cellcolor{mygreen} \textcolor{white}{$\beta$} & \cellcolor{mygreen} \textcolor{white}{$\beta$}                                              & 6 \\  
\texttt{rad}     && \cellcolor{mygreen} \textcolor{white}{$\beta$} & \cellcolor{mygreen} \textcolor{white}{$\beta$} & {} & \cellcolor{mygreen} \textcolor{white}{$\beta$} && \cellcolor{mygreen} \textcolor{white}{$\beta$} & \cellcolor{mygreen} \textcolor{white}{$\beta$} & \cellcolor{mygreen} \textcolor{white}{$\beta$}                                              & 6 \\  
\texttt{lnox}    && \cellcolor{mygreen} \textcolor{white}{$\beta$} & \cellcolor{mygreen} \textcolor{white}{$\beta$} & {} & \cellcolor{mygreen} \textcolor{white}{$\beta$} && \cellcolor{mygreen} \textcolor{white}{$\beta$} & \cellcolor{mygreen} \textcolor{white}{$\beta$} & \cellcolor{mygreen} \textcolor{white}{$\beta$}                                              & 6 \\  
\texttt{age}     && \cellcolor{mygreen} \textcolor{white}{$\beta$} & {} & {} & \cellcolor{mygreen} \textcolor{white}{$\beta$} && {} & {} & {}                                                                                                                                                                                                                              & 2 \\  
\texttt{chast}   && \cellcolor{mygreen} \textcolor{white}{$\beta$} & \cellcolor{mygreen} \textcolor{white}{$\beta$} & {} & {} && \cellcolor{mygreen} \textcolor{white}{$\beta$} & \cellcolor{mygreen} \textcolor{white}{$\beta$} & \cellcolor{mygreen} \textcolor{white}{$\beta$}                                                                                          & 5 \\  
\texttt{zn}      && {} & {} & \cellcolor{mygreen} \textcolor{white}{$\beta$} & {} && {} & {} & \cellcolor{mygreen} \textcolor{white}{$\beta$}                                                                                                                                                                                                                              & 2 \\  
\texttt{indus}   && {} & {} & {} & {} && {} & {} & {}                                                                                                                                                                                                                                                                                                                      & 0 \\  
\midrule
\texttt{ltax}    &&  \cellcolor{mygreen} \textcolor{white}{$\alpha$} & \cellcolor{mygreen} \textcolor{white}{$\alpha$} & \cellcolor{mygreen} \textcolor{white}{$\alpha$} & \cellcolor{mygreen} \textcolor{white}{$\alpha$} && {} & {} & {}  & 4 \\
\texttt{rm}      &&  {} & {} & {} & \cellcolor{mygreen} \textcolor{white}{$\alpha$} && {} & {} & {}                                                                                                                                         & 1 \\
\texttt{ldis}    &&  \cellcolor{mygreen} \textcolor{white}{$\alpha$} & \cellcolor{mygreen} \textcolor{white}{$\alpha$} & \cellcolor{mygreen} \textcolor{white}{$\alpha$} & \cellcolor{mygreen} \textcolor{white}{$\alpha$} && {} & {} & {}  & 4 \\   
\texttt{llstat}  &&  {} & {} & {} & {} && {} & {} & {}                                                                                                                                                                                      & 0 \\
\texttt{ptratio} &&  {} & {} & {} & {} && {} & {} & {}                                                                                                                                                                                      & 0 \\
\texttt{crim}    &&  {} & {} & \cellcolor{mygreen} \textcolor{white}{$\alpha$} & {} && {} & {} & {}                                                                                                                                         & 1 \\   
\texttt{rad}     &&  \cellcolor{mygreen} \textcolor{white}{$\alpha$} & \cellcolor{mygreen} \textcolor{white}{$\alpha$} & \cellcolor{mygreen} \textcolor{white}{$\alpha$} & \cellcolor{mygreen} \textcolor{white}{$\alpha$} && {} & {} & {}  & 4 \\   
\texttt{lnox}    &&  {} & {} & {} & {} && {} & {} & {}                                                                                                                                                                                      & 0 \\
\texttt{age}     &&  {} & {} & {} & {} && {} & {} & {}                                                                                                                                                                                      & 0 \\   
\texttt{chast}   &&  {} & {} & \cellcolor{mygreen} \textcolor{white}{$\alpha$} & {} && {} & {} & {}                                                                                                                                         & 1 \\   
\texttt{zn}      &&  {} & {} & \cellcolor{mygreen} \textcolor{white}{$\alpha$} & {} && {} & {} & {}                                                                                                                                         & 1 \\   
\texttt{indus}    && {} & {} & \cellcolor{mygreen} \textcolor{white}{$\alpha$} & {} && {} & {} & {}                                                                                                                                         & 1 \\    
\midrule
No. Selected    && 13        & 12        & 12        & 13        && 9         & 9         & 10       & {} \\
BIC             && -457      & -451      & -221      & -423      && -289      & -222      & -283     & {} \\
$\hat \kappa$   && 1.52      & 1.61      & 1.83      & 1.51      && 0.86      & -         &  -       & {}\\ 
Time* (mins)     && 1.42      & 2.24      & 2.80      & 559       && 0.25      & 0.002     & 1.20     & {}\\ 
\bottomrule
\multicolumn{11}{p{1\textwidth}}{\footnotesize *Intel(R) Core(TM) i7-10610U CPU @ 1.80GHz   2.30 GHz.}\\
\end{tabular}}}
 \end{table}
 
Table~\ref{tab:sgnd_hprice_vars} provides summary metrics and an overview of the variables selected by each method. The SGND-MPR and GAMLSS-STEP methods select almost the same set of covariates in both the location and the scale. In the location, all covariates apart from \texttt{zn} and \texttt{indus} are selected (and \texttt{age} is also not selected by GAMLSS-STEP but is selected by SGND-MPR), while, in the scale, only \texttt{ltax}, \texttt{ldis}, and \texttt{rad} are selected. BAMLSS also yields a similar selection, but does not select \texttt{chast} in the location and additionally selects \texttt{rm} in the scale. The GAMLSS-BOOST method selects quite a different set of covariates, and has a much higher BIC value of -221 compared to the other three distributional regression approaches, which have BIC values of -457, -451, and -423, respectively. Indeed, it would appear that GAMLSS-BOOST under-selects in the location and over-selects in the scale. The single-parameter models (SGND-SPR, ALASSO and LAD-LASSO) select the same covariates (except that LAD-LASSO selects \texttt{zn}), and this set aligns with the SGND-MPR, GAMLSS-STEP, and BAMLSS methods (except that \texttt{age} is not selected).

In the single-parameter regression models, the SGND-SPR and LAD-LASSO have more similar BIC values to each other (-289 and -283) than the ALASSO, which has a higher value (-222); this to be expected given the heavy-tailedness of the residuals in the data, as discussed in Section~\ref{sec:sgnd_introduction}. However, a more pertinent point here is that these single-parameter regression models have much higher BIC values than the distributional regression models (ignoring GAMLSS-BOOST). This demonstrates the importance of accounting for heteroscedasticity by modelling the scale in addition to the location. It is also interesting to note that the SGND-SPR method estimates quite a heavy-tailed distribution with $\hat \kappa = 0.86$. In contrast, the SGND-MPR, GAMLSS-STEP, and BAMLSS methods (which capture heteroscedasticity) yield $\hat \kappa$ values approximately between 1.5 and 1.6, implying an error distribution with tails that are heavier than normal but lighter than Laplace. Therefore, it appears that the very heavy tails suggested by the single-parameter regression analysis is, in part, an artifact of the heteroscedasticity for which they fail to account.

In terms of computational expense, the SGND-MPR method is the fastest of the distributional regression methods, taking 1.42 minutes. This is approximately 1.6 times faster than the GAMLSS-STEP approach, and two times faster than the GAMLSS-BOOST procedure. The BAMLSS method performs a three-dimensional grid search and takes over nine hours, making it very computationally demanding. For the single-parameter models, the SGND-SPR method is 4.8 times faster than the LAD-LASSO search. Note that the ALASSO method is compiled in \texttt{C} code, and so is the fastest overall.

\begin{figure}[t!]
    \centering
    \makebox{
    \begin{subfigure}[h]{\textwidth}
    \centering
    \includegraphics[width = 0.95\textwidth]{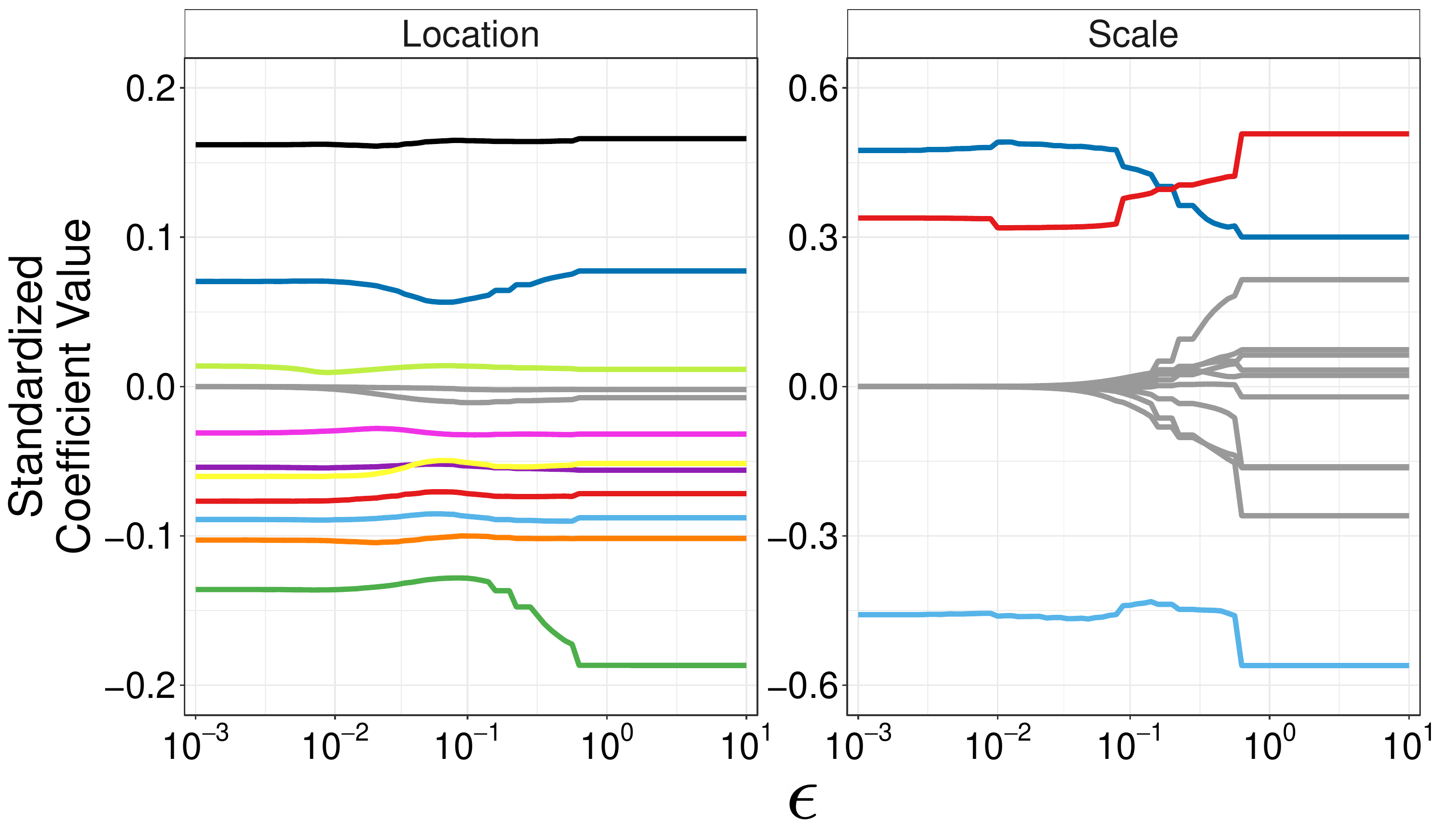}%
    \end{subfigure}}
    \makebox{
    \begin{subfigure}[h]{\textwidth}
    \centering
    \includegraphics[trim = {0.2cm 0 0.2cm 0.1cm}, clip, scale = 0.7]{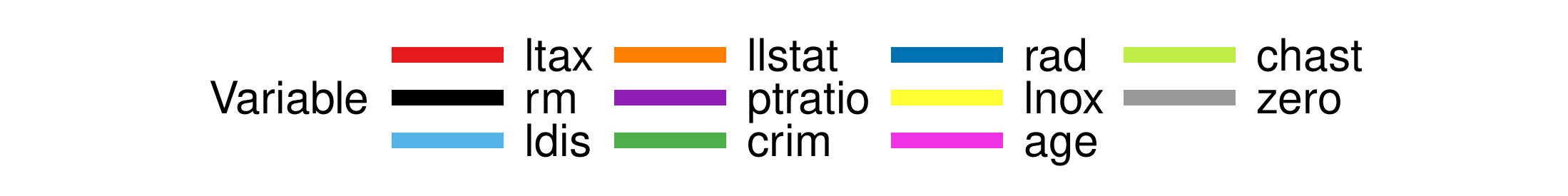}%
    \end{subfigure}}
    \caption{\label{figs:sgnd_hprice_path}Boston Housing Data: standardized coefficient values through the $\epsilon$-telescope for the location and scale components. Coloured lines indicate the selected variables, with grey lines signifying variables that are set to zero and not included in the final model.}
    \end{figure}

The standardized coefficient values with respect to the $\epsilon$-telescope of the SGND-MPR method are shown in Figure~\ref{figs:sgnd_hprice_path}. This demonstrates how the proposed method operates as $\epsilon$ moves towards zero to carry out variable selection by approximating $L_0$ penalization; indeed, by following the coefficient paths from right to left, the unimportant covariate effects tend towards zero. Two variables are eliminated from the location component, with \texttt{zn} tending to zero first, which is then followed closely by \texttt{indus} (both in grey). Although non-zero (i.e.,~selected), the \texttt{chast} and \texttt{age} effects are the weakest --- and note that BAMLSS did not select the former while GAMLSS-STEP did not select the latter. In contrast to the location, where ten variables are selected, only three variables are selected in the scale (\texttt{ltax}, \texttt{ldis} and \texttt{rad}). Interestingly, one of the last scale effects to be set to zero in the $\epsilon$-telescope is \texttt{rm}, and this variable was selected by BAMLSS.

\begin{table}[t!]
\caption{\label{tab:sgnd_dataset_estimates_hprice_mpr}Boston Housing Data: estimation metrics}
\centering
\begin{tabular}{@{}l@{~~}  r@{~}c@{~}r@{~~} c@{\qquad}  r@{~}c@{~}r@{~~} c@{\qquad} r@{}}
\toprule
{} & \multicolumn{9}{c}{SGND-MPR: $\hat \nu_0 = 0.28 \, (0.11)$}\\

\cmidrule(){2-10} 

{} & \multicolumn{2}{c}{$\hat\beta_j$} & \multicolumn{1}{c}{$\Delta\text{BIC}_\beta$} && \multicolumn{2}{c}{$\hat\alpha_j$} & \multicolumn{1}{c}{$\Delta\text{BIC}_\alpha$} && \multicolumn{1}{c}{$\Delta\text{BIC}_{\beta \alpha}$} \\
 \midrule
      \texttt{intercept} &3.67  & (0.13) &        && -8.30 & (1.97)  &      &&         \\ 
      \texttt{ltax}      &-0.19 & (0.02) & 46     && 0.85  & (0.33)  & 146  && 266     \\ 
      \texttt{rm}        &0.23  & (0.02) & 158    &&       &         &      && 158     \\ 
      \texttt{ldis}      &-0.16 & (0.03) & 57     && -0.85 & (0.18)  & 17   && 82      \\ 
      \texttt{llstat}    &-0.17 & (0.02) & 72     &&       &         &      && 72      \\ 
      \texttt{ptratio}   &-0.02 & (0.00) & 63     &&       &         &      && 63      \\ 
      \texttt{crim}      &-0.02 & (0.00) & 40     &&       &         &      && 40      \\ 
      \texttt{rad}       &0.01  & (0.00) & 12     && 0.05  & (0.02)  & 6    && 27      \\ 
      \texttt{lnox}      &-0.30 & (0.13) & 11     &&       &         &      && 11      \\ 
      \texttt{age}       &-0.00 & (0.00) & 5      &&       &         &      && 5       \\ 
      \texttt{chast}     &0.05  & (0.02) & 1      &&       &         &      && 1       \\ 
\bottomrule
\multicolumn{10}{p{0.5\textwidth}}{\footnotesize Variables not selected: \texttt{zn}, \texttt{indus}.}\\
\end{tabular}
\end{table}

For the SGND-MPR method, Table~\ref{tab:sgnd_dataset_estimates_hprice_mpr} shows the estimates and associated standard errors (in brackets) along with the $\Delta$BIC value, which is a measure of the effect of the variable and is computed as the change in BIC after removing the variable from the location ($\Delta\text{BIC}_\beta$) or scale ($\Delta\text{BIC}_\alpha$) component of the model. The overall effect of the variable (i.e.,~removal from both the location and scale components simultaneously) is denoted by $\Delta\text{BIC}_{\beta \alpha}$.\ (See the Supplementary Material for the outputs of the other methods.)\ The average number of rooms (\texttt{rm}) has a positive coefficient, indicating that more rooms in a dwelling are associated with increased house prices. Dropping this variable from the model results in an increase in the BIC of 158 units, which is substantially greater than the effect of the other variables in the location component in terms of BIC. Proximity to highways (\texttt{rad}) and houses that are situated on the banks of the Charles River (\texttt{chast}) are also linked to greater house prices, but contribute much less in terms of BIC (12 and 1 units). Communities with the following characteristics are associated with lower house prices: large property-tax rates (\texttt{ltax}), longer distances to work (\texttt{ldis}), a high percentage of ``lower status" individuals (\texttt{llstat}), high pupil teacher ratios (\texttt{ptratio}), high crime rates (\texttt{crim}), high nitric oxides concentrations (\texttt{lnox}) and the presence of older houses (\texttt{age}). Note that the top five variables in terms of BIC impact in the location component (\texttt{ltax}, \texttt{rm}, \texttt{ldis}, \texttt{llstat} and \texttt{ptratio}) appear in the location components of all of the models from Table~\ref{tab:sgnd_hprice_vars} (apart from absence of \texttt{ldis} in GAMLSS-BOOST), which is another signal of their importance.

\begin{figure}[t!]
\centering
\makebox{\includegraphics[width = 0.75\textwidth]{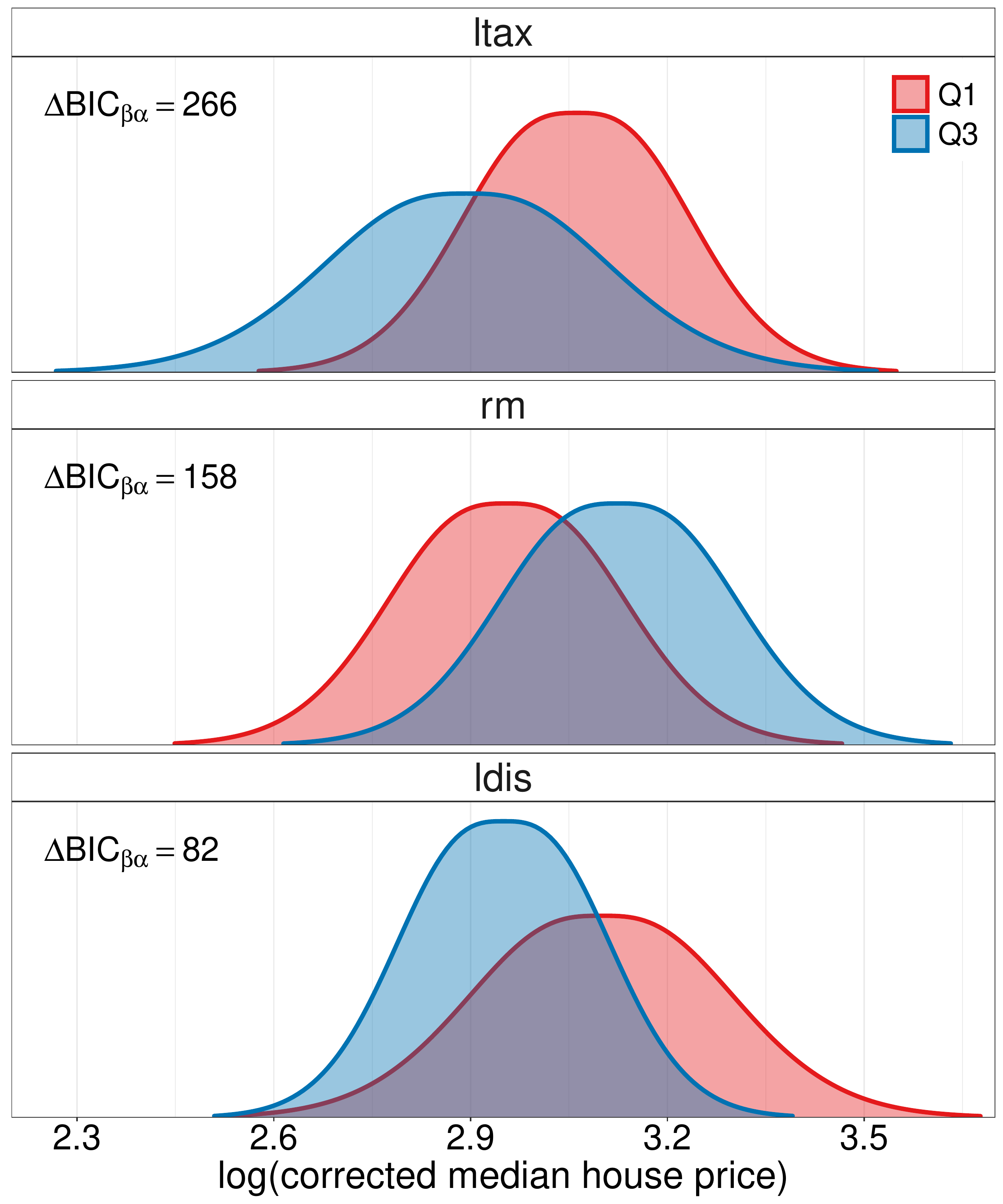}}
\caption{\label{figs:sgnd_hprice_densities}Boston Housing Data: SGND-MPR model-based conditional density curves. Keeping the covariates fixed at the median values, the red and blue densities correspond to the modification of the presented covariate as: ``low" (Q1, the first quartile) and ``high" (Q3, the third quartile).}
\end{figure}

In the case of \texttt{ltax}, although it has a negative coefficient in the location parameter, it has a positive coefficient in the scale component. Therefore, lower values of property-tax are associated with lower house prices, but also increased variability. Removal of \texttt{ltax} from the scale component results in an increase of 146 units, meaning it is the variable with the second-largest effect in terms of BIC (after \texttt{rm} in the location component with 158 units). This is also far greater than the effect of removing it from the location parameter (and all other variables in the location apart from \texttt{rm}), demonstrating the importance of modelling the scale in addition to the location in general. Removing \texttt{ltax} from both the location and scale components simultaneously results in an increase in BIC of 266 units, i.e., this is the most important covariate overall. To visualize the effect of \texttt{ltax}, Figure~\ref{figs:sgnd_hprice_densities} shows model-based conditional density curves, motivated by \citet{stadlmann21interactively}; \texttt{rm} and \texttt{ldis} are also shown and are the second and third most important variables overall. For higher values of \texttt{ltax}, there is a clear location shift to the left and greater dispersion in the associated conditional density plots. The \texttt{rm} variable only impacts the location, while the \texttt{ldis} variable has negative coefficients in both the location and scale components (reduced average house prices and variability).

\subsection{Diabetes Data}\label{sec:sgnd_real_data_analyses_diabetes}
\begin{figure}[b!]
\centering
\makebox{\includegraphics[width = \textwidth]{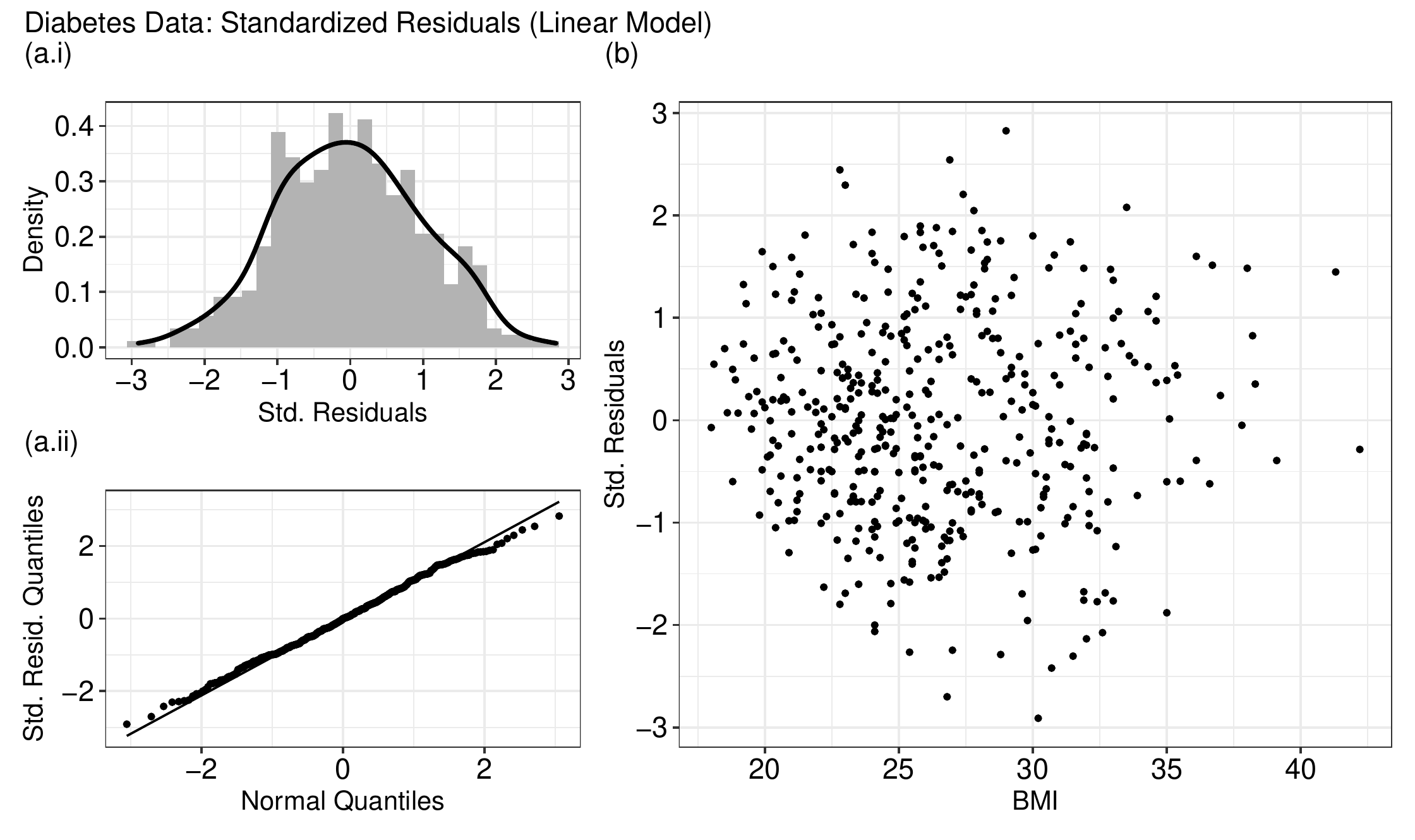}}
\caption{\label{figs:sgnd_diabets}Diabetes Data: density (a.i) and Q-Q (a.ii) plot of the standardized residuals from fitting a linear model. Standardized residuals plotted against the independent variable BMI (b).}
\end{figure}

The diabetes data come from a study where 442 diabetic patients were examined and had ten baseline variables recorded as well as the response of interest, a quantitative measure of disease progression one year after baseline. This data set is analyzed in \citet{efron04} and \citet{leng10ladlasso} and is available in the \texttt{lars} package \citep{lars_pkg}. The baseline variables are body mass index (\texttt{BMI}), average blood pressure (\texttt{BP}), sex (=1 if female, =2 if male) (\texttt{SEX}), age (\texttt{AGE}), and six blood serum measurements (\texttt{S1}, \texttt{S2}, \texttt{S3}, \texttt{S4}, \texttt{S5}, \texttt{S6}). The assumption of normality of the error distribution appears to be appropriate, as shown in Figure~\ref{figs:sgnd_diabets}(a.i) and (a.ii), which display the density and Q-Q plot of the standardized residuals from fitting a normal linear regression model. A Shapiro-Wilk test on these residuals does not provide evidence against normality (p-value = 0.6). Moreover, a bootstrapped 95\% confidence interval of the excess kurtosis contains zero [-0.52, 0.06], indicating a mesokurtic distribution, which has tails similar to that of a normal distribution. Figure~\ref{figs:sgnd_diabets}(b) shows the standardized residuals from fitting a linear model plotted against \texttt{BMI}, which highlights the possible presence of heteroscedasticity (albeit arguably not as strong as in the Boston housing data).

\begin{table}[t!]
\caption{\label{tab:sgnd_diabetes_vars}Diabetes Data: variables selected and summary metrics}
\centering
\resizebox{\textwidth}{!}{
\setlength{\tabcolsep}{4pt}{
\begin{tabular}{@{}l@{~~} c@{~} c@{~~}c@{~~}c@{~~}c@{~~} c@{~~~} c@{~~}c@{~~}c@{~~}c@{}}
\toprule
{} && \multicolumn{4}{c}{Location \& Scale} && \multicolumn{3}{c}{Location Only} & {}\\
\cmidrule(r){2-6} \cmidrule(l){7-10}
{}  &&   {\begin{tabular}{@{}c@{}}SGND \\ -MPR\end{tabular}} & {\begin{tabular}{@{}c@{}}GAMLSS \\ -STEP\end{tabular}} & {\begin{tabular}{@{}c@{}}GAMLSS \\ -BOOST\end{tabular}} & BAMLSS && {\begin{tabular}{@{}c@{}}SGND \\ -SPR\end{tabular}} & ALASSO & {\begin{tabular}{@{}c@{}}LAD \\ -LASSO\end{tabular}} & Total \\
\midrule
\texttt{BMI} && \cellcolor{mygreen} \textcolor{white}{$\beta$} & \cellcolor{mygreen} \textcolor{white}{$\beta$} & \cellcolor{mygreen} \textcolor{white}{$\beta$} & \cellcolor{mygreen} \textcolor{white}{$\beta$} && \cellcolor{mygreen} \textcolor{white}{$\beta$} & \cellcolor{mygreen} \textcolor{white}{$\beta$} & \cellcolor{mygreen} \textcolor{white}{$\beta$}  & 7 \\  
\texttt{S5}  && \cellcolor{mygreen} \textcolor{white}{$\beta$} & \cellcolor{mygreen} \textcolor{white}{$\beta$} & \cellcolor{mygreen} \textcolor{white}{$\beta$} & {} && \cellcolor{mygreen} \textcolor{white}{$\beta$} & \cellcolor{mygreen} \textcolor{white}{$\beta$} & \cellcolor{mygreen} \textcolor{white}{$\beta$}                                              & 6 \\   
\texttt{S3}  && \cellcolor{mygreen} \textcolor{white}{$\beta$} & \cellcolor{mygreen} \textcolor{white}{$\beta$} & {} & \cellcolor{mygreen} \textcolor{white}{$\beta$} && {} & {} & \cellcolor{mygreen} \textcolor{white}{$\beta$}                                                                                                                                      & 4 \\   
\texttt{BP}  && \cellcolor{mygreen} \textcolor{white}{$\beta$} & \cellcolor{mygreen} \textcolor{white}{$\beta$} & {} & \cellcolor{mygreen} \textcolor{white}{$\beta$} && \cellcolor{mygreen} \textcolor{white}{$\beta$} & \cellcolor{mygreen} \textcolor{white}{$\beta$} & \cellcolor{mygreen} \textcolor{white}{$\beta$}                                              & 6 \\       
\texttt{SEX} && \cellcolor{mygreen} \textcolor{white}{$\beta$} & \cellcolor{mygreen} \textcolor{white}{$\beta$} & {} & \cellcolor{mygreen} \textcolor{white}{$\beta$} && \cellcolor{mygreen} \textcolor{white}{$\beta$} & \cellcolor{mygreen} \textcolor{white}{$\beta$} & \cellcolor{mygreen} \textcolor{white}{$\beta$}                                              & 6 \\         
\texttt{S1}  && {} & {} & {} & \cellcolor{mygreen} \textcolor{white}{$\beta$} && \cellcolor{mygreen} \textcolor{white}{$\beta$} & \cellcolor{mygreen} \textcolor{white}{$\beta$} & \cellcolor{mygreen} \textcolor{white}{$\beta$}                                                                                         & 4 \\       
\texttt{S2}  && {} & {} & {} & \cellcolor{mygreen} \textcolor{white}{$\beta$} && \cellcolor{mygreen} \textcolor{white}{$\beta$} & \cellcolor{mygreen} \textcolor{white}{$\beta$} & {}                                                                                                                                                                                  & 3 \\                   
\texttt{S4}  && {} & {} & {} & {} && {} & \cellcolor{mygreen} \textcolor{white}{$\beta$} & {}                                                                                                                                                                                                                                                                          & 1 \\                 
\texttt{S6}  && {} & {} & {} & {} && {} & {} & \cellcolor{mygreen} \textcolor{white}{$\beta$}                                                                                                                                                                                                                                                                          & 1 \\                   
\texttt{AGE} && {} & {} & {} & {} && {} & {} & {}                                                                                                                                                                                                                                                                                                                      & 0 \\                       

\midrule

\texttt{BMI} && \cellcolor{mygreen} \textcolor{white}{$\alpha$} & \cellcolor{mygreen} \textcolor{white}{$\alpha$} & \cellcolor{mygreen} \textcolor{white}{$\alpha$} & {} && {} & {} & {}  & 3 \\  
\texttt{S5}  && {} & {} & {} & \cellcolor{mygreen} \textcolor{white}{$\alpha$} && {} & {} & {}                                                                                            & 1 \\    
\texttt{S3}  && {} & {} & {} & \cellcolor{mygreen} \textcolor{white}{$\alpha$} && {} & {} & {}                                                                                            & 1 \\                
\texttt{BP}  && {} & {} & \cellcolor{mygreen} \textcolor{white}{$\alpha$} & \cellcolor{mygreen} \textcolor{white}{$\alpha$} && {} & {} & {}                                               & 2 \\        
\texttt{SEX} && {} & {} & {} & \cellcolor{mygreen} \textcolor{white}{$\alpha$} && {} & {} & {}                                                                                            & 1 \\    
\texttt{S1}  && {} & {} & {} & \cellcolor{mygreen} \textcolor{white}{$\alpha$} && {} & {} & {}                                                                                            & 1 \\      
\texttt{S2}  && {} & {} & \cellcolor{mygreen} \textcolor{white}{$\alpha$} & \cellcolor{mygreen} \textcolor{white}{$\alpha$} && {} & {} & {}                                               & 2 \\          
\texttt{S4}  && {} & {} & {} & {} && {} & {} & {}                                                                                                                                         & 0 \\            
\texttt{S6}  && {} & {} & \cellcolor{mygreen} \textcolor{white}{$\alpha$} & {} && {} & {} & {}                                                                                            & 1 \\            
\texttt{AGE} && {} & {} & {} & {} && {} & {} & {}                                                                                                                                         & 0 \\                                                             

\midrule
No. Selected    &&  6      & 6       & 6       & 12      && 6       & 7       & 7    & {} \\
BIC             &&  4819   & 4819    & 4991    & 4886    && 4828    & 4829    & 4906 & {} \\
$\hat \kappa$   &&  2.41   & 2.65    & 4.78    & 2.26    && 2.13    &  -      &  -   & {} \\ 
Time* (mins)    &&  1.69   & 0.20    & 40.73   & 2115    && 0.39    & 0.01    & 0.66 & {} \\ 
\bottomrule
\multicolumn{11}{p{1\textwidth}}{\footnotesize *Intel(R) Core(TM) i7-10610U CPU @ 1.80GHz   2.30 GHz.}\\
\end{tabular}}}
 \end{table}

The variables selected by each method and additional summary metrics are shown in Table~\ref{tab:sgnd_diabetes_vars}. Unlike in the Boston housing data, heteroscedasticity is much less of a feature for the diabetes data since, although the lowest BIC (4819) is achieved by two distributional regression models (SGND-MPR and GAMLSS-STEP, which both select the same set of covariates in both the location and the scale), the second-lowest BIC (4828) corresponds to a single-parameter regression model (SGND-SPR). Moreover, both SGND-MPR and GAMLSS-STEP only select one scale effect (\texttt{BMI}), whereas BAMLSS appears to overestimate the extent of the heteroscedasticity with six variables selected (though not \texttt{BMI}) and also has one of the higher BIC values (4886). Similar to the Boston housing data, the GAMLSS-BOOST method has the highest BIC value (4991) among the models considered and appears to be under-selecting in the location component and over-selecting in the scale component. In the location component, \texttt{BMI} is the only covariate selected by all models; \texttt{S5} is selected by all but BAMLSS, and \texttt{BP} and \texttt{SEX} are selected by all but GAMLSS-BOOST. The models also agree that \texttt{AGE} is not important, while the \texttt{S4} and \texttt{S6} variables are each selected by just one single-parameter regression approach (ALASSO and LAD-LASSO, respectively); however, there is less agreement on the variables \texttt{S1}, \texttt{S2} and \texttt{S3}. In terms of the distributional shape, all of the methods agree on a light-tailed error distribution (in line with Figure~\ref{figs:sgnd_diabets}). GAMLSS-STEP estimates $\hat \kappa = 2.65$, which indicates an error distribution with lighter-than-normal tails, and the two SGND approaches (MPR and SPR) and BAMLSS suggest tails that are closer to that of a normal distribution ($\hat \kappa = 2.41$, $2.13$ and, $2.26$ respectively). The GAMLSS-BOOST approach suggests an error distribution with tails much lighter than the normal distribution with $\hat \kappa = 4.78$. Unsurprisingly then, the LAD-LASSO, which imposes heavy tails, has the second-highest BIC value (4906).

\begin{table}[b!]
\caption{\label{tab:sgnd_dataset_estimates_diabetes_mpr}Diabetes Data: estimation metrics}
\centering
\begin{tabular}{@{}l@{~~}  r@{~}c@{~}r@{~~} c@{\qquad}  r@{~}c@{~}r@{~~} c@{\qquad} r@{}}
\toprule
{} & \multicolumn{9}{c}{SGND-MPR: $\hat \nu_0 = 0.79 \, (0.15)$}\\
\cmidrule(){2-10} 
{} & \multicolumn{2}{c}{$\hat\beta_j$} & \multicolumn{1}{c}{$\Delta\text{BIC}_{\beta}$} && \multicolumn{2}{c}{$\hat\alpha_j$} & \multicolumn{1}{c}{$\Delta\text{BIC}_{\alpha}$} && \multicolumn{1}{c}{$\Delta\text{BIC}_{\beta \alpha}$}\\
 \midrule

\texttt{intercept} &  -183.89 & (36.18) &         && 7.04  & (0.44)  &     &&      \\
\texttt{BMI}       &  5.37    & (0.72)  & 44      && 0.06  & (0.02)  & 7   && 65   \\
\texttt{S5}        &  40.77   & (5.77)  & 43      &&       &         &     && 43   \\
\texttt{S3}        &  -1.03   & (0.24)  & 13      &&       &         &     && 13   \\
\texttt{BP}        &  0.92    & (0.22)  & 12      &&       &         &     && 12   \\
\texttt{SEX}       &  -20.46  & (5.89)  & 6       &&       &         &     && 6    \\
\bottomrule
\multicolumn{10}{p{0.5\textwidth}}{\footnotesize Variables not selected: \texttt{S1}, \texttt{S2}, \texttt{S4}, \texttt{S6}, \texttt{AGE}.}\\
\end{tabular}
\end{table}

\begin{figure}[h!]
    \centering
    \makebox{
    \begin{subfigure}[h]{\textwidth}
    \centering
    \includegraphics[width = 0.95\textwidth]{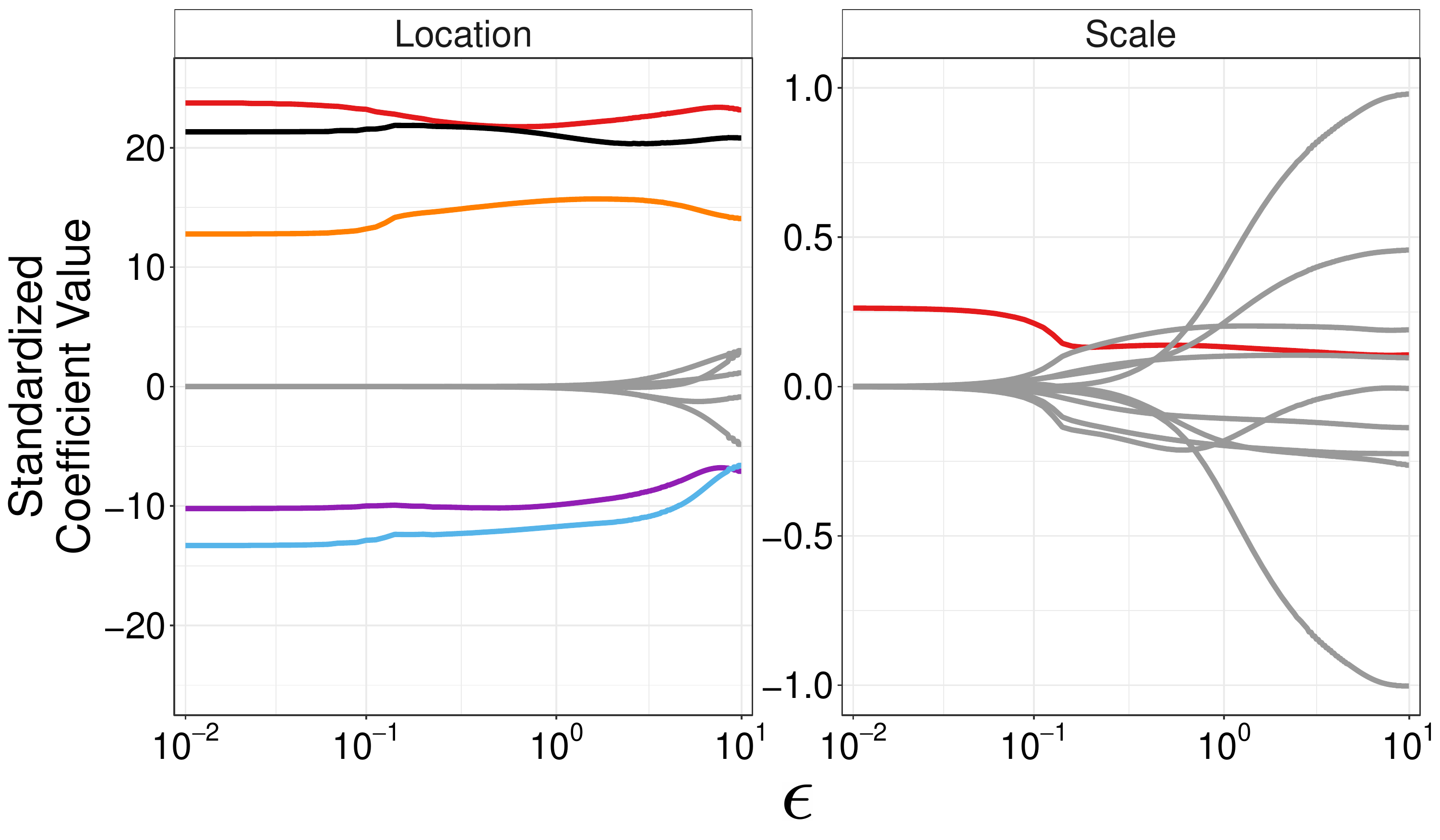}%
    \end{subfigure}}
    \makebox{
    \begin{subfigure}[h]{\textwidth}
    \centering
    \includegraphics[trim = {0.2cm 0 0.2cm 0.1cm}, clip, scale = 0.7]{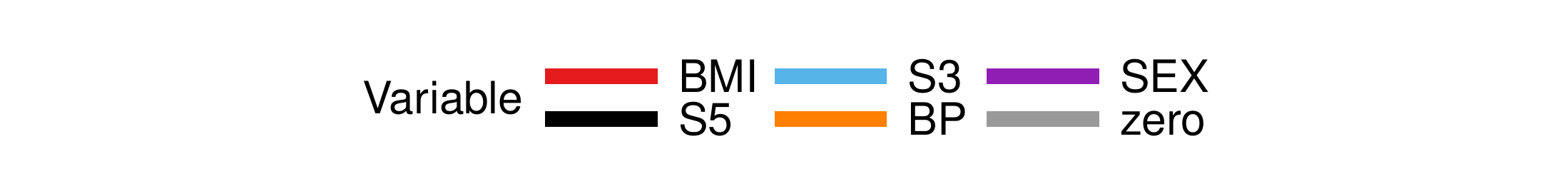}%
    \end{subfigure}}
    \caption{\label{fig:sgnd_diabetes_path}Diabetes Data: standardized coefficient values through the $\epsilon$-telescope for the location and scale components. Coloured lines indicate the selected variables, with grey lines signifying variables that are set to zero and not included in the final model.}
    \end{figure}

\begin{figure}[t!]
\centering
\makebox{\includegraphics[width = 0.75\textwidth]{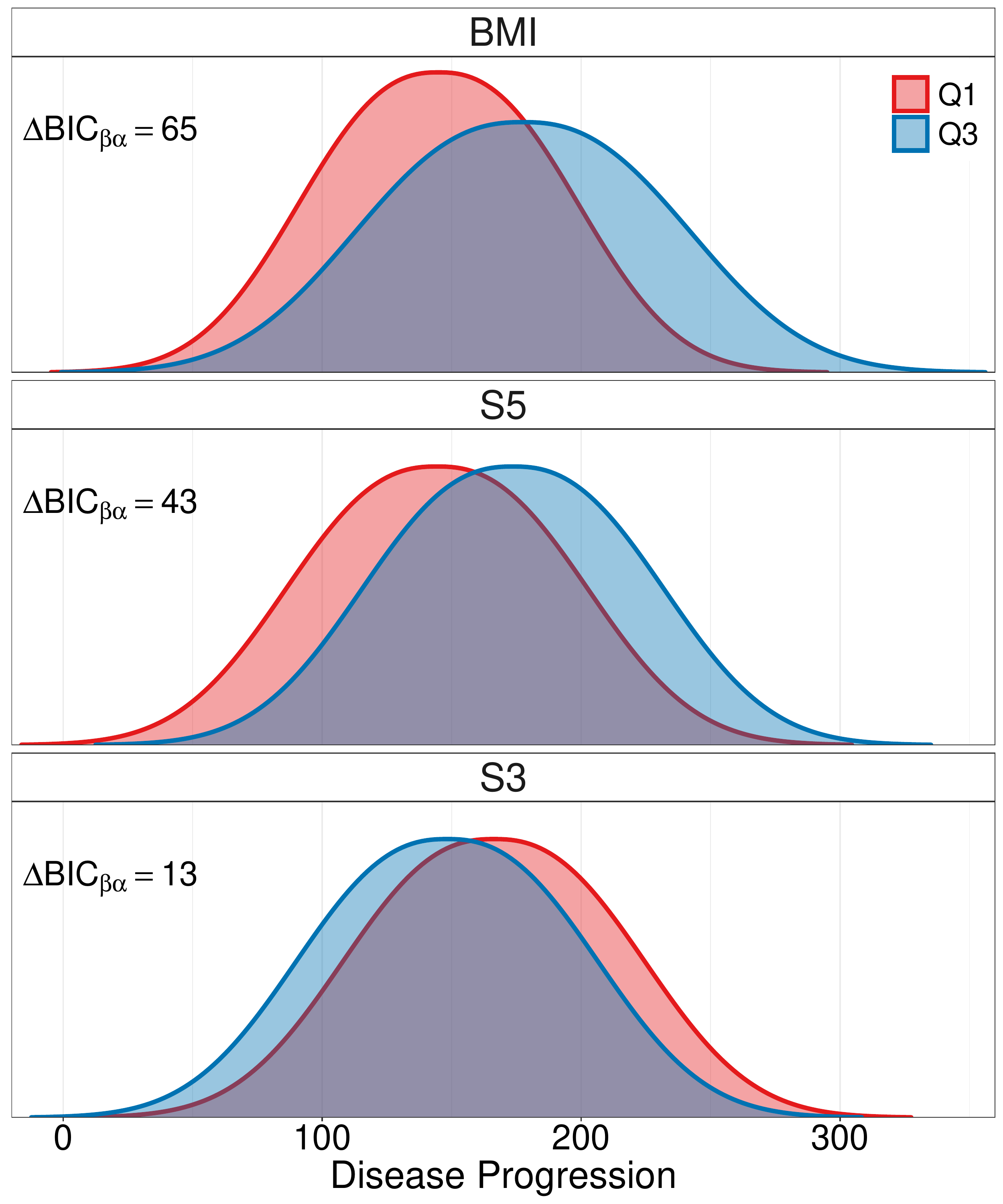}}
\caption{\label{figs:sgnd_diabetes_densities}Diabetes Data: SGND-MPR model-based conditional density curves. Keeping the covariates fixed at the median values, the red and blue densities correspond to the modification of the presented covariate as: ``low" (Q1, the first quartile) and ``high" (Q3, the third quartile).}
\end{figure} 

Table~\ref{tab:sgnd_dataset_estimates_diabetes_mpr} displays the estimates (with coefficient paths visualized in Figure~\ref{fig:sgnd_diabetes_path}), standard errors and $\Delta$BIC values for the SGND-MPR method (estimation metrics for the other methods are available in the Supplementary Material). For the location component, the top two variables in terms of BIC impact are \texttt{BMI} and \texttt{S5} as previously suggested by Table~\ref{tab:sgnd_diabetes_vars}. Both of these variables are associated with increased disease progression, as is \texttt{BP}, while \texttt{S3} and \texttt{SEX} are associated with reduced disease progression. In the case of \texttt{SEX}, this translates to progression being lower in males. The only scale effect identified here is \texttt{BMI}, such that larger BMI values are associated with increased variability. However, this scale effect ($\Delta\text{BIC}_\alpha = 7$) is less important than the top four location effects, i.e., heteroscedasticity is not a strong feature of this dataset. Nevertheless, when modelled simultaneously in the location and scale, \texttt{BMI} is the most important variable ($\Delta\text{BIC}_{\beta \alpha}$ = 65). The effect of the top three variables in terms of overall BIC impact (\texttt{BMI}, \texttt{S5} and \texttt{S3}) are also visualized in Figure~\ref{figs:sgnd_diabetes_densities} using model-based conditional density curves.

\section{Discussion} \label{sec:sgnd_discussion}
Our proposed SGND-SIC method provides a flexible framework that (i) allows for the error distribution to move between (or beyond) robust and classical normal linear regression, (ii) accounts for heteroscedasticity by taking a distributional regression approach where the scale is modelled on covariates in addition to the location, and (iii) carries out automatic variable selection in both distributional regression components (location and scale) in a computationally feasible manner; we have also implemented the approach in the \texttt{smoothic} package in \texttt{R} \citep{smoothic_package}. The presence of outliers and heteroscedasticity can distort parameter estimates and make standard inferences unreliable. Through the estimation of the shape parameter in the smooth generalized normal distribution, our proposed method can move smoothly between the normal distribution and the heavier-tailed Laplace distribution. Therefore, one does not have to make an outright decision between classical and robust regression, as both are covered. Moreover, allowing the scale to depend on covariates reduces the risk of information loss, as examination of regression effects beyond the mean is key to identifying important dynamics in the data. Currently, variable selection is somewhat underdeveloped in the distributional regression setting, with existing penalization methods being particularly computationally intensive in this setting. We avoid the typical multidimensional grid search for optimal tuning parameter values, as we know from the outset that $\lambda = \log(n)$ in the BIC case. Our smooth objective function can be optimized directly using standard gradient based techniques and performs comparably to existing procedures as illustrated in the real data analyses.

Through extensive simulation studies, we have shown that our procedure performs favorably in terms of variable selection and parameter inference over the full range of $\kappa$ values that are tested. We have also demonstrated the importance of accounting for both heavy tails and heteroscedasticity when they exist. Although standard error breakdown appears to occur when the combination of heavy tails and heteroscedasticity is observed in smaller samples, the procedure performs well asymptotically --- and in smaller sample sizes when only one of heavy tails or heteroscedasticity is observed. Nevertheless, it will be useful to investigate alternative standard error calculations in our future work. Moreover, it is also of interest to extend our approach to non-symmetric distributions and other parameter types beyond location and scale; indeed, there may be merit in some settings in modelling higher-order shape parameters, such as $\kappa$ in the generalized normal distribution.

\section*{Acknowledgements}
This work was carried out within the Confirm Smart Manufacturing Research Centre (https://confirm.ie/) funded by Science Foundation Ireland (grant number: 16/RC/3918).

\bibliographystyle{apalike}
\bibliography{bibfile}

\begin{thebibliography}{}

\bibitem[Amato et~al., 2021]{amato21penalised}
Amato, U., Antoniadis, A., De~Feis, I., and Gijbels, I. (2021).
\newblock Penalised robust estimators for sparse and high-dimensional linear
  models.
\newblock {\em Statistical Methods \& Applications}, 30(1):1--48.

\bibitem[Amin et~al., 2015]{amin15scad}
Amin, M., Song, L., Thorlie, M.~A., and Wang, X. (2015).
\newblock {SCAD}-penalized quantile regression for high-dimensional data
  analysis and variable selection.
\newblock {\em Statistica Neerlandica}, 69(3):212--235.

\bibitem[Avella~Medina and Ronchetti, 2015]{avella15robustreview}
Avella~Medina, M. and Ronchetti, E. (2015).
\newblock Robust statistics: a selective overview and new directions.
\newblock {\em Wiley Interdisciplinary Reviews: Computational Statistics},
  7(6):372--393.

\bibitem[Bassett~Jr and Koenker, 1978]{bassett78asymptotic}
Bassett~Jr, G. and Koenker, R. (1978).
\newblock Asymptotic theory of least absolute error regression.
\newblock {\em Journal of the American Statistical Association},
  73(363):618--622.

\bibitem[Belloni and Chernozhukov, 2011]{belloni11}
Belloni, A. and Chernozhukov, V. (2011).
\newblock l1-penalized quantile regression in high-dimensional sparse models.
\newblock {\em The Annals of Statistics}, 39(1):82--130.

\bibitem[Box and Tiao, 1973]{box73pe}
Box, G.~E. and Tiao, G.~C. (1973).
\newblock {\em Bayesian Inference in Statistical Analysis}.
\newblock MA: Addison-Wesley Publishing Co.

\bibitem[Burke et~al., 2020]{burke20}
Burke, K., Jones, M., and Noufaily, A. (2020).
\newblock A flexible parametric modelling framework for survival analysis.
\newblock {\em Journal of the Royal Statistical Society: Series C (Applied
  Statistics)}, 69(2):429--457.

\bibitem[Cox and Reid, 1987]{cox87}
Cox, D.~R. and Reid, N. (1987).
\newblock Parameter orthogonality and approximate conditional inference.
\newblock {\em Journal of the Royal Statistical Society: Series B
  (Methodological)}, 49(1):1--18.

\bibitem[Croux et~al., 2004]{croux04robust}
Croux, C., Dhaene, G., and Hoorelbeke, D. (2004).
\newblock Robust standard errors for robust estimators.
\newblock {\em CES-Discussion paper series (DPS) 03.16}, pages 1--20.

\bibitem[Dielman, 2005]{dielman05least}
Dielman, T.~E. (2005).
\newblock Least absolute value regression: recent contributions.
\newblock {\em Journal of statistical computation and simulation},
  75(4):263--286.

\bibitem[Efron et~al., 2004]{efron04}
Efron, B., Hastie, T., Johnstone, I., Tibshirani, R., et~al. (2004).
\newblock Least angle regression.
\newblock {\em The Annals of Statistics}, 32(2):407--499.

\bibitem[Fan et~al., 2014]{fan14robustadaptive}
Fan, J., Fan, Y., and Barut, E. (2014).
\newblock Adaptive robust variable selection.
\newblock {\em Annals of Statistics}, 42(1):324.

\bibitem[Fan et~al., 2017]{fan17estimation}
Fan, J., Li, Q., and Wang, Y. (2017).
\newblock Estimation of high dimensional mean regression in the absence of
  symmetry and light tail assumptions.
\newblock {\em Journal of the Royal Statistical Society: Series B (Statistical
  Methodology)}, 79(1):247--265.

\bibitem[Fan and Li, 2001]{fan01}
Fan, J. and Li, R. (2001).
\newblock Variable selection via nonconcave penalized likelihood and its oracle
  properties.
\newblock {\em Journal of the American statistical Association},
  96(456):1348--1360.

\bibitem[Fan and Li, 2002]{fan02}
Fan, J. and Li, R. (2002).
\newblock Variable selection for {Cox's} proportional hazards model and frailty
  model.
\newblock {\em Annals of Statistics}, pages 74--99.

\bibitem[Filzmoser and Nordhausen, 2021]{filzmoser21robust}
Filzmoser, P. and Nordhausen, K. (2021).
\newblock Robust linear regression for high-dimensional data: An overview.
\newblock {\em Wiley Interdisciplinary Reviews: Computational Statistics},
  13(4):e1524.

\bibitem[Friedman et~al., 2010]{friedman10}
Friedman, J., Hastie, T., and Tibshirani, R. (2010).
\newblock Regularization paths for generalized linear models via coordinate
  descent.
\newblock {\em Journal of statistical software}, 33(1):1.

\bibitem[Groll et~al., 2019]{groll19}
Groll, A., Hambuckers, J., Kneib, T., and Umlauf, N. (2019).
\newblock {LASSO}-type penalization in the framework of generalized additive
  models for location, scale and shape.
\newblock {\em Computational Statistics \& Data Analysis}, 140:59--73.

\bibitem[Hambuckers et~al., 2018]{hambuckers18understanding}
Hambuckers, J., Groll, A., and Kneib, T. (2018).
\newblock Understanding the economic determinants of the severity of
  operational losses: A regularized generalized pareto regression approach.
\newblock {\em Journal of Applied Econometrics}, 33(6):898--935.

\bibitem[Hanin, 2021]{hanin21cavalier}
Hanin, L. (2021).
\newblock Cavalier use of inferential statistics is a major source of false and
  irreproducible scientific findings.
\newblock {\em Mathematics}, 9(6):603.

\bibitem[Harrison~Jr and Rubinfeld, 1978]{harrison78hprice2}
Harrison~Jr, D. and Rubinfeld, D.~L. (1978).
\newblock Hedonic housing prices and the demand for clean air.
\newblock {\em Journal of environmental economics and management},
  5(1):81--102.

\bibitem[Harvey, 1976]{harvey76}
Harvey, A.~C. (1976).
\newblock Estimating regression models with multiplicative heteroscedasticity.
\newblock {\em Econometrica: Journal of the Econometric Society}, pages
  461--465.

\bibitem[Hastie and Efron, 2022]{lars_pkg}
Hastie, T. and Efron, B. (2022).
\newblock {\em lars: Least Angle Regression, Lasso and Forward Stagewise}.
\newblock {R} package version 1.3.

\bibitem[Hofner et~al., 2016]{gamboostLSS}
Hofner, B., Mayr, A., and Schmid, M. (2016).
\newblock {gamboostLSS}: An {R} package for model building and variable
  selection in the {GAMLSS} framework.
\newblock {\em Journal of Statistical Software}, 74(1):1–31.

\bibitem[Jiang et~al., 2021]{jiang21outlier}
Jiang, Y., Wang, Y., Zhang, J., Xie, B., Liao, J., and Liao, W. (2021).
\newblock Outlier detection and robust variable selection via the penalized
  weighted {LAD-LASSO} method.
\newblock {\em Journal of Applied Statistics}, 48(2):234--246.

\bibitem[Jones and Anaya-Izquierdo, 2011]{jones11orthog}
Jones, M. and Anaya-Izquierdo, K. (2011).
\newblock On parameter orthogonality in symmetric and skew models.
\newblock {\em Journal of Statistical Planning and Inference}, 141(2):758--770.

\bibitem[Kneib, 2013]{kneib13beyond}
Kneib, T. (2013).
\newblock Beyond mean regression.
\newblock {\em Statistical Modelling}, 13(4):275--303.

\bibitem[Kneib et~al., 2021]{kneib21rage}
Kneib, T., Silbersdorff, A., and S{\"a}fken, B. (2021).
\newblock Rage against the mean--a review of distributional regression
  approaches.
\newblock {\em Econometrics and Statistics}.

\bibitem[Koenker and Bassett~Jr, 1978]{koenker78quantiles}
Koenker, R. and Bassett~Jr, G. (1978).
\newblock Regression quantiles.
\newblock {\em Econometrica: journal of the Econometric Society}, pages 33--50.

\bibitem[Koenker and Hallock, 2001]{quantregbook}
Koenker, R. and Hallock, K.~F. (2001).
\newblock Quantile regression.
\newblock {\em Journal of economic perspectives}, 15(4):143--156.

\bibitem[Kuhn et~al., 2013]{kuhn13applied}
Kuhn, M., Johnson, K., et~al. (2013).
\newblock {\em Applied predictive modeling}, volume~26.
\newblock Springer.

\bibitem[Leisch and Dimitriadou, 2021]{mlbench_pkg}
Leisch, F. and Dimitriadou, E. (2021).
\newblock {\em mlbench: Machine Learning Benchmark Problems}.
\newblock R package version 2.1-3.

\bibitem[Leng, 2010]{leng10ladlasso}
Leng, C. (2010).
\newblock Variable selection and coefficient estimation via regularized rank
  regression.
\newblock {\em Statistica Sinica}, pages 167--181.

\bibitem[Li et~al., 2015]{li15flarepaper}
Li, X., Zhao, T., Yuan, X., and Liu, H. (2015).
\newblock The flare package for high dimensional linear regression and
  precision matrix estimation in {R}.
\newblock {\em Journal of Machine Learning Research}.

\bibitem[Maronna et~al., 2019]{maronna19robustbook}
Maronna, R.~A., Martin, R.~D., Yohai, V.~J., and Salibi{\'a}n-Barrera, M.
  (2019).
\newblock {\em Robust statistics: theory and methods (with R)}.
\newblock John Wiley \& Sons.

\bibitem[Mayr et~al., 2012]{mayr12}
Mayr, A., Fenske, N., Hofner, B., Kneib, T., and Schmid, M. (2012).
\newblock Generalized additive models for location, scale and shape for high
  dimensional data—a flexible approach based on boosting.
\newblock {\em Journal of the Royal Statistical Society: Series C (Applied
  Statistics)}, 61(3):403--427.

\bibitem[Nadarajah, 2005]{nadarajah05generalized}
Nadarajah, S. (2005).
\newblock A generalized normal distribution.
\newblock {\em Journal of Applied statistics}, 32(7):685--694.

\bibitem[Nelson, 1991]{nelson91conditional}
Nelson, D.~B. (1991).
\newblock Conditional heteroskedasticity in asset returns: A new approach.
\newblock {\em Econometrica: Journal of the Econometric Society}, pages
  347--370.

\bibitem[O'Neill and Burke, 2021]{oneill2021smoothicarxiv}
O'Neill, M. and Burke, K. (2021).
\newblock Variable selection using a smooth information criterion for
  multi-parameter regression models.
\newblock {\em arXiv preprint arXiv:2110.02643}.

\bibitem[O'Neill and Burke, 2022]{smoothic_package}
O'Neill, M. and Burke, K. (2022).
\newblock {\em smoothic: Variable selection using a smooth information
  criterion}.
\newblock {R} package version 1.0.0.

\bibitem[Osorio and Wolodzko, 2022]{L1pack}
Osorio, F. and Wolodzko, T. (2022).
\newblock {\em Routines for L1 estimation}.
\newblock R package version 0.41-2.

\bibitem[{R Core Team}, 2022]{citer}
{R Core Team} (2022).
\newblock {\em {R}: A Language and Environment for Statistical Computing}.
\newblock R Foundation for Statistical Computing, Vienna, Austria.

\bibitem[Ramires et~al., 2021]{ramires21gamlssstep}
Ramires, T.~G., Nakamura, L.~R., Righetto, A.~J., Pescim, R.~R., Mazucheli, J.,
  Rigby, R.~A., and Stasinopoulos, D.~M. (2021).
\newblock Validation of stepwise-based procedure in {GAMLSS}.
\newblock {\em Journal of Data Science}, 19(1):96--110.

\bibitem[Reid et~al., 2016]{reid16}
Reid, S., Tibshirani, R., and Friedman, J. (2016).
\newblock A study of error variance estimation in lasso regression.
\newblock {\em Statistica Sinica}, pages 35--67.

\bibitem[Rigby and Stasinopoulos, 2005]{rigby05}
Rigby, R.~A. and Stasinopoulos, D.~M. (2005).
\newblock Generalized additive models for location, scale and shape.
\newblock {\em Journal of the Royal Statistical Society: Series C (Applied
  Statistics)}, 54(3):507--554.

\bibitem[Ronchetti, 2020]{ronchetti20accurate}
Ronchetti, E. (2020).
\newblock Accurate and robust inference.
\newblock {\em Econometrics and Statistics}, 14:74--88.

\bibitem[Rousseeuw and Croux, 1993]{rousseeuw93robustscale}
Rousseeuw, P.~J. and Croux, C. (1993).
\newblock Alternatives to the median absolute deviation.
\newblock {\em Journal of the American Statistical association},
  88(424):1273--1283.

\bibitem[Rutemiller and Bowers, 1968]{rutemiller68}
Rutemiller, H.~C. and Bowers, D.~A. (1968).
\newblock Estimation in a heteroscedastic regression model.
\newblock {\em Journal of the American Statistical Association},
  63(322):552--557.

\bibitem[Salibian-Barrera and Zamar, 2002]{salibian02}
Salibian-Barrera, M. and Zamar, R.~H. (2002).
\newblock Bootstrapping robust estimates of regression.
\newblock {\em Annals of Statistics}, pages 556--582.

\bibitem[Simpson et~al., 1992]{simpson92se}
Simpson, D.~G., Ruppert, D., and Carroll, R.~J. (1992).
\newblock On one-step {GM} estimates and stability of inferences in linear
  regression.
\newblock {\em Journal of the American Statistical Association},
  87(418):439--450.

\bibitem[Singh, 1998]{singh98breakdown}
Singh, K. (1998).
\newblock Breakdown theory for bootstrap quantiles.
\newblock {\em The Annals of Statistics}, 26(5):1719--1732.

\bibitem[Smucler and Yohai, 2017]{smucler17robust}
Smucler, E. and Yohai, V.~J. (2017).
\newblock Robust and sparse estimators for linear regression models.
\newblock {\em Computational Statistics \& Data Analysis}, 111:116--130.

\bibitem[Stadlmann and Kneib, 2021]{stadlmann21interactively}
Stadlmann, S. and Kneib, T. (2021).
\newblock Interactively visualizing distributional regression models with
  distreg.vis.
\newblock {\em Statistical Modelling}, page 1471082X211007308.

\bibitem[Stasinopoulos and Rigby, 2007]{stas07}
Stasinopoulos, D.~M. and Rigby, R.~A. (2007).
\newblock Generalized additive models for location scale and shape ({GAMLSS})
  in {R}.
\newblock {\em Journal of Statistical Software}, 23(7):1–46.

\bibitem[Stasinopoulos et~al., 2018]{stas18gamlss}
Stasinopoulos, M.~D., Rigby, R.~A., and Bastiani, F.~D. (2018).
\newblock {GAMLSS}: a distributional regression approach.
\newblock {\em Statistical Modelling}, 18(3-4):248--273.

\bibitem[Stasinopoulos et~al., 2017]{stasinopoulos17gamlssbook}
Stasinopoulos, M.~D., Rigby, R.~A., Heller, G.~Z., Voudouris, V., and
  De~Bastiani, F. (2017).
\newblock {\em Flexible regression and smoothing: using {GAMLSS} in {R}}.
\newblock CRC Press.

\bibitem[Tibshirani, 1996]{tibshirani96}
Tibshirani, R. (1996).
\newblock Regression shrinkage and selection via the lasso.
\newblock {\em Journal of the Royal Statistical Society. Series B
  (Methodological)}, pages 267--288.

\bibitem[Tibshirani, 1997]{tibshirani97}
Tibshirani, R. (1997).
\newblock The lasso method for variable selection in the {Cox} model.
\newblock {\em Statistics in Medicine}, 16(4):385--395.

\bibitem[Todorov and Filzmoser, 2009]{robustbasepaper}
Todorov, V. and Filzmoser, P. (2009).
\newblock An object-oriented framework for robust multivariate analysis.
\newblock {\em Journal of Statistical Software}, 32(3):1--47.

\bibitem[Umlauf et~al., 2018]{umlauf18bamlsspaper}
Umlauf, N., Klein, N., and Zeileis, A. (2018).
\newblock {BAMLSS}: Bayesian additive models for location, scale, and shape
  (and beyond).
\newblock {\em Journal of Computational and Graphical Statistics},
  27(3):612--627.

\bibitem[Venables and Ripley, 2002]{massbook}
Venables, W.~N. and Ripley, B.~D. (2002).
\newblock {\em Modern Applied Statistics with {S}}.
\newblock Springer, New York, fourth edition.
\newblock ISBN 0-387-95457-0.

\bibitem[Wang et~al., 2007]{wang07robust}
Wang, H., Li, G., and Jiang, G. (2007).
\newblock Robust regression shrinkage and consistent variable selection through
  the {LAD-Lasso}.
\newblock {\em Journal of Business \& Economic Statistics}, 25(3):347--355.

\bibitem[Wu and Liu, 2009]{wu09variable}
Wu, Y. and Liu, Y. (2009).
\newblock Variable selection in quantile regression.
\newblock {\em Statistica Sinica}, pages 801--817.

\bibitem[Yu, 2013]{yu13stability}
Yu, B. (2013).
\newblock Stability.
\newblock {\em Bernoulli}, 19(4):1484--1500.

\end{thebibliography}

\clearpage
\appendix

\section{Analytical Derivatives} 
\subsection{Derivatives of the Normalizing Constant}\label{app:sgnd_derivatives_ctilde}
The normalizing constant in the ``smooth generalized normal distribution" (SGND) is
\begin{equation*}
    \tilde{c}_{\tau}(\kappa)=\frac{1}{\int_{-\infty}^{\infty} e^{-\tilde{g}(z)} dz },
\end{equation*}
where $\tilde{g}(z)=\left(\sqrt{z^{2}+\tau^{2}}-\tau\right)^{\kappa}$ (see main paper). In order to compute the derivatives of $\ell_{i}^{\text{SIC}}$, the derivatives of $\log(\tilde{c}_{\tau}(\kappa))$ are required and are given below, with $\tilde a_i = \sqrt{y_i^2 + \tau^2} - \tau$:
    \begin{align*} 
            	\begin{split}	
            	\frac{\partial}{\partial \nu}\log(\tilde{c}_{\tau}(\kappa)) =& \frac{\frac{\partial}{\partial \nu}\tilde{c}_{\tau}(\kappa)}{\tilde{c}_{\tau}(\kappa)},\\
            	\frac{\partial}{\partial \nu}\tilde{c}_{\tau}(\kappa)=&\frac{-\int_{-\infty}^{\infty} \frac{\partial}{\partial \nu} e^{-\tilde{g}(z)} dz}{\left(\int_{-\infty}^{\infty}  e^{-\tilde{g}(z)} dz \right)^2},\\
            	\frac{\partial}{\partial \nu} e^{-\tilde{g}(z)}=&-\tilde a_i^{\kappa}e^{-\tilde a_i^{\kappa}}\log(\tilde a_i)(\kappa - \kappa_\text{min}),\\
            	\frac{\partial^2}{\partial \nu^2}\log(\tilde{c}_{\tau}(\kappa)) =& \frac{\tilde{c}_{\tau}(\kappa) \frac{\partial^2}{\partial \nu^2}\tilde{c}_{\tau}(\kappa)-(\frac{\partial}{\partial \nu}\tilde{c}_{\tau}(\kappa))^2}{\tilde{c}_{\tau}(\kappa)^2},\\
            	\frac{\partial^2}{\partial \nu^2}\tilde{c}_{\tau}(\kappa)=&\frac{\left(\int_{-\infty}^{\infty}  e^{-\tilde{g}(z)}dz \right) \left(-\int_{-\infty}^{\infty} \frac{\partial^2}{\partial \nu^2} e^{-\tilde{g}(z)} dz \right) + 2\left(\int_{-\infty}^{\infty} \frac{\partial}{\partial \nu} e^{-\tilde{g}(z)} dz \right)^2}{\left(\int_{-\infty}^{\infty}  e^{-\tilde{g}(z)} dz \right)^3},\\
            	\frac{\partial^2}{\partial \nu^2} e^{-\tilde{g}(z)}=&-\tilde a_i^\kappa e^{-\tilde a_i^{\kappa}} \log^2(\tilde a_i)(\kappa-\kappa_\text{min})^2-\\
             &\tilde a_i^{\kappa} e^{-\tilde a_i^\kappa} \log(\tilde a_i)(\kappa - \kappa_\text{min})\left(1-\tilde a_i^\kappa \log(\tilde a_i)(\kappa - \kappa_\text{min})\right).
            	\end{split}
    \end{align*}

\subsection{Second Derivatives of the Log-Likelihood Function}\label{app:sgnd_derivatives_second}
The second derivatives of $\ell_{i}^{\text{SIC}}$ are required for the diagonal weight matrices and are given below. These are used in the hessian matrix (see main paper).
\begin{align*} 
    \begin{split}
    W_{\beta, i} =&\frac{s_i^{-2}\kappa a_{i}^{\kappa-1}}{a_{i}+\tau}+\frac{s_{i}^{-4}\kappa(y_i-\mu_i)^{2}a_{i}^{\kappa}\left(\sqrt{a_{i}^2+2(\tau a_{i}+\frac{\tau^2}{2})}(\kappa-1)a_i^{-2}-a_i^{-1}\right)}{{(a_{i}^2+2(\tau a_{i}+\frac{\tau^2}{2}))^{\frac{3}{2}}}},\\
    W_{\alpha, i} =& \frac{s_i^{-2}\kappa(y_i-\mu_i)^{2}a_{i}^{\kappa -1}}{2(a_{i} + \tau)} - \frac{s_i^{-4}\kappa(y_i-\mu_i)^{4}a_{i}^{\kappa-1}}{{4(a_{i}^2+2(\tau a_{i}+\frac{\tau^2}{2}))^{\frac{3}{2}}}}+\frac{s_i^{-4}(\kappa-1)\kappa(y_i-\mu_i)^{4}a_{i}^{\kappa-2}}{4(a_{i}^2+2(\tau a_{i}+\frac{\tau^2}{2}))},\\
    W_{\nu, i} =& -\frac{\partial^2}{\partial \nu^2}\log(\tilde{c}_{\tau}(\kappa))-\left((\kappa - \kappa_\text{min}) \log(a_{i})a_{i}^{\kappa}\right)\left((\kappa - \kappa_\text{min}) \log(a_{i})-1\right),\\
    W_{\beta \alpha, i} =& \frac{s_i^{-2}\kappa(y_i-\mu_i)a_{i}^{\kappa-1}}{a_{i}+\tau} - \frac{s_i^{-4}\kappa(y_i-\mu_i)^{3}a_{i}^{\kappa-1}}{2(a_{i}^2+2(\tau a_{i}+\frac{\tau^2}{2}))^{\frac{3}{2}}}+\frac{s_i^{-4}(\kappa-1)\kappa(y_i-\mu_i)^{3}a_{i}^{\kappa-2}}{2(a_{i}^2+2(\tau a_{i}+\frac{\tau^2}{2}))},\\
    W_{\beta \nu, i} =& \frac{-(\kappa - \kappa_\text{min})s_i^{-2}(y_i-\mu_i)a_{i}^{\kappa-1}(1+\kappa\log (a_{i}))}{a_{i}+\tau},\\
    W_{\alpha \nu, i} =& \frac{-(\kappa - \kappa_\text{min})s_i^{-2}(y_i-\mu_i)^{2}a_{i}^{\kappa-1}(1+\kappa\log(a_{i}))}{2(a_{i}+\tau)}.
    \end{split} 
\end{align*}

\clearpage
\section{Overview of ALASSO and LAD-LASSO methods} \label{app:sgnd_additional_code}

We use the \texttt{glmnet} package \citep{friedman10} to apply the ALASSO method, where we select the value of the tuning parameter by minimizing the BIC. Note that this procedure is not available in \texttt{glmnet}, which only offers cross-validation minimization. Following \citet{reid16}, we estimate the scale parameter using the formula 
\begin{equation*}
    \hat{s}^2=\sum^{n}_{i=1}\frac{(y_i-{x_{i}}^{T}{\hat{\beta}})^2}{n-{h}},
\end{equation*}
where $h$ is the number of non-zero elements in the adaptive LASSO estimate $\hat \beta$. The ALASSO method can be applied using the following code.
\begin{lstlisting}[language=R1]
# ALASSO ----------------------------------
library(glmnet)
# Calculate adaptive weights using least squares estimates
coef_init <- coef(lm(lcmedv ~ ., data = bostonhouseprice2)) # linear model
param_weight_vec <- 1/abs(coef_init[-1]) # -1 to remove intercept

# Provides results over a range of lambdas
fit <- glmnet(x = x, # matrix of covariates
              y = lcmedv, # response vector
              penalty.factor = param_weight_vec) # adaptive weights
# Use the resulting estimates to calculate BIC & get optimal lambda
\end{lstlisting}

The \texttt{flare} package \citep{li15flarepaper} is used to apply the LAD-LASSO method. This package does not provide automatic tuning parameter selection, and, therefore, like with the ALASSO method, we have developed a tuning parameter selection procedure based on minimizing the BIC. The \texttt{flare} package provides $\hat \beta$ coefficient estimates at a given value of the tuning parameter, from which we then estimate the scale parameter via $\hat s = 1.4826 \times \text{median}\lvert y_i-x^T_i \hat \beta \rvert$; this is commonly used as a robust estimator of scale \citep{rousseeuw93robustscale, maronna19robustbook}. We obtain the LAD-LASSO estimates at a given value of the tuning parameter using the following code.
\begin{lstlisting}[language=R1]
# LAD-LASSO -------------------------------
library(flare)
# Assume the optimal lambda (opt_lambda) was found using BIC search
fit <- slim(X = x, # matrix of covariates
            Y = lcmedv, # response vector
            lambda = opt_lambda, # optimal lambda value
            method = "lq", # Lq LASSO
            q = 1) # LAD-LASSO
\end{lstlisting}
\newpage
\vspace*{0.5em}
\begin{spacing}{1.5}
\begin{center}
    {\LARGE Supplementary Material for ``Robust Distributional\\ Regression with Automatic Variable Selection"} \par
    \vskip 1.5em
       {\large   Meadhbh O'Neill \qquad \qquad Kevin Burke}
\end{center}
\end{spacing}

\counterwithin{figure}{section}
\counterwithin{table}{section}
\renewcommand\thefigure{\thesection\arabic{figure}}
\renewcommand\thetable{\thesection\arabic{table}}

\section{Simulation Results: $\boldsymbol{\tau = 0.05}$} \label{app:sgnd_smalltau}
This section contains the results of a simulation study similar to Section~3 of the main paper, but where $\tau = 0.05$. The method performs well, but experiences standard error breakdown in the heavy-tailed setting with $\kappa = 1$, resulting in poor coverage.
\begin{itemize}
    \item Table~\ref{tab:sgnd_var_selection_smalltau} is analogous to Table~2 of the main paper, but showing the model selection metrics when $\tau = 0.05$.
    \item Table~\ref{tab:sgnd_parameter_inference_smalltau} is analogous to Table~3 of the main paper, but showing the estimation and inference metrics when $\tau = 0.05$.
\end{itemize}

\begin{table}[h!]
\caption{Simulation results: model selection metrics} 
\label{tab:sgnd_var_selection_smalltau}
\centering
\resizebox{\textwidth}{!}{
\begin{tabular}{@{}c@{~~~}l@{~~~}   c@{~~}c@{~~}c@{~~}   c@{~~}   c@{~~}c@{~~}c@{~~}   c@{~~}   c@{~~}c@{~~}c@{~~}   c@{~~}   c@{~~}c@{~~}c@{}}
  \toprule
 {} & {} & \multicolumn{3}{c}{$\kappa = 1$} && \multicolumn{3}{c}{$\kappa = 1.33$} && \multicolumn{3}{c}{$\kappa = 1.67$} && \multicolumn{3}{c}{$\kappa = 2$} \\
 \cmidrule(r){3-5} \cmidrule(r){7-9} \cmidrule(r){11-13} \cmidrule(){15-17}
 {} & $n$ & C(6) & PT & MSE && C(6) & PT & MSE && C(6) & PT & MSE && C(6) & PT & MSE\\ 
 \midrule
 {$\beta$}                & 500   & 5.76 & 0.80 & 0.01 && 5.87 & 0.88 & 0.01 && 5.89 & 0.90 & 0.01 && 5.90 & 0.91 & 0.01 \\ 
 {}                       & 1000  & 5.86 & 0.87 & 0.01 && 5.91 & 0.92 & 0.00 && 5.93 & 0.94 & 0.00 && 5.94 & 0.94 & 0.00 \\ 
 {}                       & 5000  & 5.96 & 0.96 & 0.00 && 5.97 & 0.97 & 0.00 && 5.97 & 0.97 & 0.00 && 5.97 & 0.97 & 0.00 \\[0.2cm]
 
 {$\alpha$}                & 500   & 5.90 & 0.89 & 0.15 && 5.89 & 0.90 & 0.08 && 5.89 & 0.90 & 0.06 && 5.89 & 0.90 & 0.04 \\ 
 {}                        & 1000  & 5.94 & 0.94 & 0.06 && 5.94 & 0.94 & 0.04 && 5.94 & 0.94 & 0.03 && 5.94 & 0.94 & 0.02 \\ 
 {}                        & 5000  & 5.99 & 0.99 & 0.01 && 5.99 & 0.99 & 0.01 && 5.99 & 0.99 & 0.00 && 5.99 & 0.99 & 0.00 \\
   \bottomrule
   \multicolumn{17}{p{0.9\textwidth}}{\footnotesize C, average correct zeros; PT, the probability of choosing the true model; MSE, the average mean squared error.}\\
\end{tabular}}
\end{table}

\begin{table}[h!]
\caption{Simulation results: estimation and inference metrics}
\label{tab:sgnd_parameter_inference_smalltau}
\begin{subtable}{\textwidth}
    \centering
    \subcaption{}
    \vspace{-0.1cm}
\begin{tabular}{@{}c@{~~}c@{~~~} c@{~~}c@{~~}c@{~~}c@{~~} c@{~~~} c@{~~}c@{~~}c@{~~}c@{~~} c@{~~~} c@{~~}c@{~~}c@{~~}c@{}}
\toprule
\multicolumn{2}{l}{$\kappa=1$} & \multicolumn{4}{c}{$n = 500$} && \multicolumn{4}{c}{$n = 1000$} && \multicolumn{4}{c}{$n = 5000$} \\
\cmidrule(r){3-6} \cmidrule(r){8-11} \cmidrule(){13-16}
{} & $\theta$ & $\hat{\theta}$ & SE & SEE & CP && $\hat{\theta}$ & SE & SEE & CP && $\hat{\theta}$ & SE & SEE & CP \\
  \midrule
  $\beta_{0}$   & 0.00  & -0.00 & 0.07 & 0.03 & 0.43 && -0.00 & 0.05 & 0.03 & 0.55 && -0.00 & 0.02 & 0.02 & 0.80 \\
  $\beta_{1}$   & 1.00  & 1.00  & 0.05 & 0.03 & 0.50 && 1.00  & 0.03 & 0.02 & 0.62 && 1.00  & 0.01 & 0.01 & 0.90 \\ 
  $\beta_{2}$   & 0.50  & 0.50  & 0.04 & 0.02 & 0.40 && 0.50  & 0.03 & 0.01 & 0.50 && 0.50  & 0.01 & 0.01 & 0.78 \\ 
  $\beta_{3}$   & 0.50  & 0.50  & 0.03 & 0.02 & 0.51 && 0.50  & 0.02 & 0.02 & 0.65 && 0.50  & 0.01 & 0.01 & 0.90 \\ 
  $\beta_{4}$   & 1.00  & 1.00  & 0.04 & 0.02 & 0.45 && 1.00  & 0.02 & 0.01 & 0.59 && 1.00  & 0.01 & 0.01 & 0.85 \\ 
  $\beta_{5}$   & 0.50  & 0.50  & 0.04 & 0.02 & 0.56 && 0.50  & 0.02 & 0.02 & 0.66 && 0.50  & 0.01 & 0.01 & 0.91 \\ 
  $\beta_{6}$   & 1.00  & 1.00  & 0.04 & 0.02 & 0.49 && 1.00  & 0.03 & 0.02 & 0.60 && 1.00  & 0.01 & 0.01 & 0.84 \\[0.2cm]
  $\alpha_{0}$  & 0.00  & -0.08 & 0.32 & 0.27 & 0.91 && -0.03 & 0.20 & 0.18 & 0.93 && -0.01 & 0.08 & 0.08 & 0.93 \\
  $\alpha_{1}$  & 0.50  & 0.51  & 0.10 & 0.09 & 0.94 && 0.50  & 0.07 & 0.06 & 0.94 && 0.50  & 0.03 & 0.03 & 0.94 \\ 
  $\alpha_{2}$  & 1.00  & 1.02  & 0.09 & 0.09 & 0.92 && 1.01  & 0.07 & 0.06 & 0.94 && 1.00  & 0.03 & 0.03 & 0.95 \\ 
  $\alpha_{3}$  & 0.50  & 0.51  & 0.11 & 0.10 & 0.94 && 0.51  & 0.08 & 0.07 & 0.94 && 0.50  & 0.03 & 0.03 & 0.95 \\ 
  $\alpha_{4}$  & 1.00  & 1.02  & 0.10 & 0.09 & 0.92 && 1.01  & 0.06 & 0.06 & 0.94 && 1.00  & 0.03 & 0.03 & 0.95 \\ 
  $\alpha_{7}$  & 0.50  & 0.51  & 0.10 & 0.09 & 0.95 && 0.51  & 0.06 & 0.06 & 0.95 && 0.50  & 0.03 & 0.03 & 0.95 \\ 
  $\alpha_{8}$  & 1.00  & 1.02  & 0.10 & 0.09 & 0.93 && 1.01  & 0.06 & 0.06 & 0.95 && 1.00  & 0.03 & 0.03 & 0.96 \\[0.2cm]
  $\nu_{0}$     & -0.22 & -0.23 & 0.12 & 0.10 & 0.89 && -0.22 & 0.08 & 0.07 & 0.91 && -0.22 & 0.03 & 0.03 & 0.92 \\
  \bottomrule
  \end{tabular}
\end{subtable}

\bigskip

\begin{subtable}{\textwidth}
    \centering
    \subcaption{}
    \vspace{-0.1cm}
\begin{tabular}{@{}c@{~~}c@{~~~} c@{~~}c@{~~}c@{~~}c@{~~} c@{~~~} c@{~~}c@{~~}c@{~~}c@{~~} c@{~~~} c@{~~}c@{~~}c@{~~}c@{}}
\toprule
\multicolumn{2}{l}{$\kappa=1.33$} & \multicolumn{4}{c}{$n = 500$} && \multicolumn{4}{c}{$n = 1000$} && \multicolumn{4}{c}{$n = 5000$} \\
\cmidrule(r){3-6} \cmidrule(r){8-11} \cmidrule(){13-16}
{} & $\theta$ & $\hat{\theta}$ & SE & SEE & CP && $\hat{\theta}$ & SE & SEE & CP && $\hat{\theta}$ & SE & SEE & CP \\
  \midrule
  $\beta_{0}$   & 0.00 & -0.00 & 0.06 & 0.05 & 0.80 && -0.00 & 0.04 & 0.03 & 0.87 && 0.00  & 0.02 & 0.02 & 0.93 \\
  $\beta_{1}$   & 1.00 & 1.00  & 0.04 & 0.04 & 0.87 && 1.00  & 0.03 & 0.03 & 0.90 && 1.00  & 0.01 & 0.01 & 0.93 \\ 
  $\beta_{2}$   & 0.50 & 0.50  & 0.03 & 0.03 & 0.78 && 0.50  & 0.02 & 0.02 & 0.85 && 0.50  & 0.01 & 0.01 & 0.92 \\ 
  $\beta_{3}$   & 0.50 & 0.50  & 0.03 & 0.02 & 0.87 && 0.50  & 0.02 & 0.02 & 0.91 && 0.50  & 0.01 & 0.01 & 0.94 \\ 
  $\beta_{4}$   & 1.00 & 1.00  & 0.03 & 0.02 & 0.80 && 1.00  & 0.02 & 0.02 & 0.89 && 1.00  & 0.01 & 0.01 & 0.94 \\ 
  $\beta_{5}$   & 0.50 & 0.50  & 0.03 & 0.02 & 0.87 && 0.50  & 0.02 & 0.02 & 0.90 && 0.50  & 0.01 & 0.01 & 0.95 \\ 
  $\beta_{6}$   & 1.00 & 1.00  & 0.03 & 0.03 & 0.80 && 1.00  & 0.02 & 0.02 & 0.87 && 1.00  & 0.01 & 0.01 & 0.93 \\[0.2cm]
  $\alpha_{0}$  & 0.00 & -0.01 & 0.22 & 0.20 & 0.92 && -0.01 & 0.14 & 0.14 & 0.95 && -0.00 & 0.06 & 0.06 & 0.94 \\
  $\alpha_{1}$  & 0.50 & 0.51  & 0.08 & 0.08 & 0.94 && 0.50  & 0.06 & 0.05 & 0.94 && 0.50  & 0.02 & 0.02 & 0.94 \\ 
  $\alpha_{2}$  & 1.00 & 1.02  & 0.08 & 0.08 & 0.93 && 1.01  & 0.06 & 0.05 & 0.94 && 1.00  & 0.02 & 0.02 & 0.95 \\ 
  $\alpha_{3}$  & 0.50 & 0.51  & 0.09 & 0.09 & 0.94 && 0.51  & 0.07 & 0.06 & 0.93 && 0.50  & 0.03 & 0.03 & 0.95 \\ 
  $\alpha_{4}$  & 1.00 & 1.02  & 0.08 & 0.08 & 0.92 && 1.01  & 0.06 & 0.05 & 0.94 && 1.00  & 0.02 & 0.02 & 0.95 \\ 
  $\alpha_{7}$  & 0.50 & 0.50  & 0.08 & 0.08 & 0.95 && 0.50  & 0.06 & 0.05 & 0.95 && 0.50  & 0.02 & 0.02 & 0.95 \\ 
  $\alpha_{8}$  & 1.00 & 1.02  & 0.08 & 0.08 & 0.93 && 1.01  & 0.06 & 0.06 & 0.95 && 1.00  & 0.02 & 0.02 & 0.96 \\[0.2cm]
  $\nu_{0}$     & 0.13 & 0.15  & 0.12 & 0.11 & 0.92 && 0.14  & 0.08 & 0.07 & 0.92 && 0.13  & 0.03 & 0.03 & 0.91 \\
  \bottomrule
  \end{tabular}
\end{subtable}
\end{table}


\begin{table}[h!]
\ContinuedFloat
\begin{subtable}{\textwidth}
    \centering
    \subcaption{}
\begin{tabular}{@{}c@{~~}c@{~~~} c@{~~}c@{~~}c@{~~}c@{~~} c@{~~~} c@{~~}c@{~~}c@{~~}c@{~~} c@{~~~} c@{~~}c@{~~}c@{~~}c@{}}
\toprule
\multicolumn{2}{l}{$\kappa=1.67$} & \multicolumn{4}{c}{$n = 500$} && \multicolumn{4}{c}{$n = 1000$} && \multicolumn{4}{c}{$n = 5000$} \\
\cmidrule(r){3-6} \cmidrule(r){8-11} \cmidrule(){13-16}
{} & $\theta$ & $\hat{\theta}$ & SE & SEE & CP && $\hat{\theta}$ & SE & SEE & CP && $\hat{\theta}$ & SE & SEE & CP \\
  \midrule
  $\beta_{0}$   & 0.00 & -0.00 & 0.05 & 0.04 & 0.90 && -0.00 & 0.03 & 0.03 & 0.93 && 0.00  & 0.01 & 0.01 & 0.94 \\
  $\beta_{1}$   & 1.00 & 1.00  & 0.04 & 0.03 & 0.92 && 1.00  & 0.02 & 0.02 & 0.95 && 1.00  & 0.01 & 0.01 & 0.94 \\ 
  $\beta_{2}$   & 0.50 & 0.50  & 0.03 & 0.03 & 0.92 && 0.50  & 0.02 & 0.02 & 0.93 && 0.50  & 0.01 & 0.01 & 0.94 \\ 
  $\beta_{3}$   & 0.50 & 0.50  & 0.02 & 0.02 & 0.94 && 0.50  & 0.02 & 0.02 & 0.95 && 0.50  & 0.01 & 0.01 & 0.94 \\ 
  $\beta_{4}$   & 1.00 & 1.00  & 0.02 & 0.02 & 0.91 && 1.00  & 0.02 & 0.02 & 0.94 && 1.00  & 0.01 & 0.01 & 0.95 \\ 
  $\beta_{5}$   & 0.50 & 0.50  & 0.02 & 0.02 & 0.94 && 0.50  & 0.02 & 0.02 & 0.94 && 0.50  & 0.01 & 0.01 & 0.96 \\ 
  $\beta_{6}$   & 1.00 & 1.00  & 0.03 & 0.03 & 0.91 && 1.00  & 0.02 & 0.02 & 0.93 && 1.00  & 0.01 & 0.01 & 0.94 \\[0.2cm]
  $\alpha_{0}$  & 0.00 & -0.00 & 0.17 & 0.16 & 0.93 && -0.00 & 0.12 & 0.11 & 0.94 && -0.00 & 0.05 & 0.05 & 0.95 \\
  $\alpha_{1}$  & 0.50 & 0.50  & 0.07 & 0.07 & 0.94 && 0.50  & 0.05 & 0.05 & 0.94 && 0.50  & 0.02 & 0.02 & 0.94 \\ 
  $\alpha_{2}$  & 1.00 & 1.02  & 0.07 & 0.07 & 0.93 && 1.01  & 0.05 & 0.05 & 0.93 && 1.00  & 0.02 & 0.02 & 0.95 \\ 
  $\alpha_{3}$  & 0.50 & 0.51  & 0.08 & 0.08 & 0.94 && 0.50  & 0.06 & 0.06 & 0.94 && 0.50  & 0.02 & 0.03 & 0.95 \\ 
  $\alpha_{4}$  & 1.00 & 1.02  & 0.07 & 0.07 & 0.92 && 1.01  & 0.05 & 0.05 & 0.94 && 1.00  & 0.02 & 0.02 & 0.95 \\ 
  $\alpha_{7}$  & 0.50 & 0.50  & 0.07 & 0.07 & 0.94 && 0.50  & 0.05 & 0.05 & 0.95 && 0.50  & 0.02 & 0.02 & 0.95 \\ 
  $\alpha_{8}$  & 1.00 & 1.01  & 0.07 & 0.07 & 0.93 && 1.01  & 0.05 & 0.05 & 0.95 && 1.00  & 0.02 & 0.02 & 0.96 \\[0.2cm]
  $\nu_{0}$     & 0.38 & 0.43  & 0.12 & 0.11 & 0.92 && 0.40  & 0.08 & 0.08 & 0.94 && 0.38  & 0.03 & 0.03 & 0.95 \\
  \bottomrule
  \end{tabular}
\end{subtable}

\bigskip

\begin{subtable}{\textwidth}
    \centering
    \subcaption{}
\begin{tabular}{@{}c@{~~}c@{~~~} c@{~~}c@{~~}c@{~~}c@{~~} c@{~~~} c@{~~}c@{~~}c@{~~}c@{~~} c@{~~~} c@{~~}c@{~~}c@{~~}c@{}}
\toprule
\multicolumn{2}{l}{$\kappa=2$} & \multicolumn{4}{c}{$n = 500$} && \multicolumn{4}{c}{$n = 1000$} && \multicolumn{4}{c}{$n = 5000$} \\
\cmidrule(r){3-6} \cmidrule(r){8-11} \cmidrule(){13-16}
{} & $\theta$ & $\hat{\theta}$ & SE & SEE & CP && $\hat{\theta}$ & SE & SEE & CP && $\hat{\theta}$ & SE & SEE & CP \\
  \midrule
  $\beta_{0}$   & 0.00 & -0.00 & 0.04 & 0.04 & 0.91 && -0.00 & 0.03 & 0.03 & 0.94 && -0.00 & 0.01 & 0.01 & 0.95 \\
  $\beta_{1}$   & 1.00 & 1.00  & 0.03 & 0.03 & 0.93 && 1.00  & 0.02 & 0.02 & 0.95 && 1.00  & 0.01 & 0.01 & 0.95 \\ 
  $\beta_{2}$   & 0.50 & 0.50  & 0.03 & 0.02 & 0.92 && 0.50  & 0.02 & 0.02 & 0.94 && 0.50  & 0.01 & 0.01 & 0.93 \\ 
  $\beta_{3}$   & 0.50 & 0.50  & 0.02 & 0.02 & 0.94 && 0.50  & 0.01 & 0.01 & 0.95 && 0.50  & 0.01 & 0.01 & 0.94 \\ 
  $\beta_{4}$   & 1.00 & 1.00  & 0.02 & 0.02 & 0.91 && 1.00  & 0.02 & 0.02 & 0.94 && 1.00  & 0.01 & 0.01 & 0.95 \\ 
  $\beta_{5}$   & 0.50 & 0.50  & 0.02 & 0.02 & 0.94 && 0.50  & 0.01 & 0.01 & 0.95 && 0.50  & 0.01 & 0.01 & 0.96 \\ 
  $\beta_{6}$   & 1.00 & 1.00  & 0.03 & 0.02 & 0.90 && 1.00  & 0.02 & 0.02 & 0.93 && 1.00  & 0.01 & 0.01 & 0.94 \\[0.2cm]
  $\alpha_{0}$  & 0.00 & -0.00 & 0.15 & 0.14 & 0.93 && -0.00 & 0.10 & 0.10 & 0.94 && -0.00 & 0.04 & 0.04 & 0.95 \\
  $\alpha_{1}$  & 0.50 & 0.50  & 0.07 & 0.06 & 0.94 && 0.50  & 0.05 & 0.04 & 0.94 && 0.50  & 0.02 & 0.02 & 0.94 \\ 
  $\alpha_{2}$  & 1.00 & 1.02  & 0.07 & 0.06 & 0.92 && 1.01  & 0.05 & 0.04 & 0.93 && 1.00  & 0.02 & 0.02 & 0.95 \\ 
  $\alpha_{3}$  & 0.50 & 0.51  & 0.07 & 0.07 & 0.94 && 0.51  & 0.05 & 0.05 & 0.94 && 0.50  & 0.02 & 0.02 & 0.95 \\ 
  $\alpha_{4}$  & 1.00 & 1.02  & 0.07 & 0.06 & 0.92 && 1.01  & 0.04 & 0.04 & 0.94 && 1.00  & 0.02 & 0.02 & 0.95 \\ 
  $\alpha_{7}$  & 0.50 & 0.50  & 0.07 & 0.06 & 0.94 && 0.50  & 0.05 & 0.04 & 0.95 && 0.50  & 0.02 & 0.02 & 0.95 \\ 
  $\alpha_{8}$  & 1.00 & 1.01  & 0.07 & 0.06 & 0.92 && 1.01  & 0.05 & 0.04 & 0.94 && 1.00  & 0.02 & 0.02 & 0.96 \\[0.2cm]
  $\nu_{0}$     & 0.59 & 0.64  & 0.13 & 0.12 & 0.91 && 0.61  & 0.08 & 0.08 & 0.93 && 0.59  & 0.03 & 0.03 & 0.94 \\
  \bottomrule
  \multicolumn{16}{p{0.86\textwidth}}{\footnotesize SE, standard deviation of estimates over 1000 replications; SEE, average of estimated standard errors over 1000 replications; CP, the empirical coverage probability of a nominal 95\% confidence interval.}\\
  \end{tabular}
\end{subtable}
\end{table}

\clearpage
\section{Simulation Results: Homoscedastic Setting} \label{app:sgnd_kappa1_homo}
This section displays additional simulation results for data simulated from the SGND with constant variance for $\kappa = 1$ and $\tau = 0.05$. In the homoscedastic setting, the issue of standard error breakdown is not experienced, and the method performs well in the heavy-tailed setting for small $\tau$.
\begin{itemize}
    \item Table~\ref{tab:sgnd_kappa1_homo_model_selection} is analogous to Table~2 of the main paper, but showing the model selection metrics in the homoscedastic setting with $\tau = 0.05$.
    \item Table~\ref{tab:sgnd_kappa1_homo_param_inference} is analogous to Table~3 of the main paper, but showing the estimation and inference metrics in the homoscedastic setting with $\tau = 0.05$.
\end{itemize}

\begin{table}[h!]
\caption{\label{tab:sgnd_kappa1_homo_model_selection}Simulation results: model selection metrics}
\centering
\begin{tabular}{@{}l@{~~~~} c@{~~}c@{~~}c@{~~}   c@{~~}   c@{~~}c@{~~}c@{~~}   c@{~~}   c@{~~}c@{~~}c@{}}
\toprule
{} & \multicolumn{3}{c}{$n = 500$} && \multicolumn{3}{c}{$n = 1000$} && \multicolumn{3}{c}{$n = 5000$} \\
\cmidrule(r){2-4} \cmidrule(r){6-8} \cmidrule(){10-12}
{$\kappa$} & C(6) & PT & MSE && C(6) & PT & MSE && C(6) & PT & MSE \\
  \midrule
  1 & 5.86 & 0.87 & 0.02 && 5.93 & 0.93 & 0.01 && 5.98 & 0.98 & 0.00 \\
  \bottomrule
  \multicolumn{12}{p{0.65\textwidth}}{\footnotesize C, average correct zeros; PT, the probability of choosing the true model; MSE, the average mean squared error.}\\
 \end{tabular}
 \end{table}

\begin{table}[h!]
\caption{Simulation results: estimation and inference metrics}
\label{tab:sgnd_kappa1_homo_param_inference}
\resizebox{\textwidth}{!}{
\begin{tabular}{@{}l@{~~} c@{~~}c@{~~}  c@{~~}c@{~~}c@{~~}c@{~~}  c@{~~}  c@{~~}c@{~~}c@{~~}c@{~~}  c@{~~}  c@{~~}c@{~~}c@{~~}c@{}}
\toprule
{} & {} & {} & \multicolumn{4}{c}{$n = 500$} && \multicolumn{4}{c}{$n = 1000$} && \multicolumn{4}{c}{$n = 5000$} \\
\cmidrule(r){4-7} \cmidrule(r){9-12} \cmidrule(){14-17}
{$\kappa$} & {} & $\theta$ & $\hat{\theta}$ & SE & SEE & CP && $\hat{\theta}$ & SE & SEE & CP && $\hat{\theta}$ & SE & SEE & CP \\
  \midrule
  1 & $\beta_{0}$   & 0.00    & 0.00  & 0.08 & 0.07 & 0.89 && 0.00  & 0.05 & 0.05 & 0.93 && -0.00 & 0.02 & 0.02 & 0.95 \\
    & $\beta_{1}$   & 1.00    & 1.00  & 0.05 & 0.05 & 0.88 && 1.00  & 0.04 & 0.03 & 0.91 && 1.00  & 0.01 & 0.01 & 0.94 \\
    & $\beta_{2}$   & 0.50    & 0.50  & 0.06 & 0.05 & 0.91 && 0.50  & 0.04 & 0.04 & 0.93 && 0.50  & 0.02 & 0.02 & 0.95 \\
    & $\beta_{3}$   & 0.50    & 0.50  & 0.06 & 0.05 & 0.91 && 0.50  & 0.04 & 0.04 & 0.94 && 0.50  & 0.02 & 0.02 & 0.94 \\
    & $\beta_{4}$   & 1.00    & 1.00  & 0.05 & 0.05 & 0.90 && 1.00  & 0.04 & 0.03 & 0.92 && 1.00  & 0.01 & 0.01 & 0.94 \\
    & $\beta_{5}$   & 0.50    & 0.50  & 0.05 & 0.05 & 0.91 && 0.50  & 0.03 & 0.03 & 0.94 && 0.50  & 0.01 & 0.01 & 0.96 \\
    & $\beta_{6}$   & 1.00    & 1.00  & 0.06 & 0.05 & 0.90 && 1.00  & 0.04 & 0.04 & 0.91 && 1.00  & 0.02 & 0.02 & 0.94 \\[0.2cm]
    & $\alpha_{0}$  & 0.00    & -0.10 & 0.29 & 0.26 & 0.93 && -0.04 & 0.18 & 0.17 & 0.94 && -0.01 & 0.08 & 0.07 & 0.93 \\[0.2cm]
    & $\nu_{0}$     & -0.22   & -0.25 & 0.12 & 0.10 & 0.91 && -0.23 & 0.07 & 0.07 & 0.91 && -0.23 & 0.03 & 0.03 & 0.92 \\
    \bottomrule
    \multicolumn{17}{p{0.9\textwidth}}{\footnotesize SE, standard deviation of estimates over 1000 replications; SEE, average of estimated standard errors over 1000 replications; CP, the empirical coverage probability of a nominal 95\% confidence interval.}\\
  \end{tabular}}
\end{table}

\clearpage
\section{Simulation Results: Additional Results} \label{app:sgnd_simulation}
This section contains additional results of the simulation study carried out in Section~3 of the main paper.

\subsection{Simulation MPR Parameter Inference} \label{app:sgnd_simulation_inference}
 Table~\ref{tab:sgnd_parameter_inference_full} contains the full results of Table~3 of the main paper.

\begin{table}[h!]
\caption{Simulation results: estimation and inference metrics}
\label{tab:sgnd_parameter_inference_full}
\begin{subtable}{\textwidth}
    \centering
    \subcaption{}
    \vspace{-0.1cm}
\begin{tabular}{@{}c@{~~}c@{~~~} c@{~~}c@{~~}c@{~~}c@{~~} c@{~~~} c@{~~}c@{~~}c@{~~}c@{~~} c@{~~~} c@{~~}c@{~~}c@{~~}c@{}}
\toprule
\multicolumn{2}{l}{$\kappa=1$} & \multicolumn{4}{c}{$n = 500$} && \multicolumn{4}{c}{$n = 1000$} && \multicolumn{4}{c}{$n = 5000$} \\
\cmidrule(r){3-6} \cmidrule(r){8-11} \cmidrule(){13-16}
{} & $\theta$ & $\hat{\theta}$ & SE & SEE & CP && $\hat{\theta}$ & SE & SEE & CP && $\hat{\theta}$ & SE & SEE & CP \\
  \midrule
  $\beta_{0}$   & 0.00  & -0.00 & 0.07 & 0.05 & 0.71 && -0.00 & 0.05 & 0.04 & 0.79 && -0.00 & 0.02 & 0.02 & 0.90 \\
  $\beta_{1}$   & 1.00  & 1.00  & 0.05 & 0.04 & 0.74 && 1.00  & 0.04 & 0.03 & 0.83 && 1.00  & 0.01 & 0.01 & 0.92 \\ 
  $\beta_{2}$   & 0.50  & 0.50  & 0.04 & 0.03 & 0.66 && 0.50  & 0.03 & 0.02 & 0.74 && 0.50  & 0.01 & 0.01 & 0.86 \\ 
  $\beta_{3}$   & 0.50  & 0.50  & 0.03 & 0.03 & 0.77 && 0.50  & 0.02 & 0.02 & 0.85 && 0.50  & 0.01 & 0.01 & 0.93 \\ 
  $\beta_{4}$   & 1.00  & 1.00  & 0.04 & 0.03 & 0.66 && 1.00  & 0.02 & 0.02 & 0.80 && 1.00  & 0.01 & 0.01 & 0.90 \\ 
  $\beta_{5}$   & 0.50  & 0.50  & 0.04 & 0.03 & 0.76 && 0.50  & 0.02 & 0.02 & 0.85 && 0.50  & 0.01 & 0.01 & 0.94 \\ 
  $\beta_{6}$   & 1.00  & 1.00  & 0.04 & 0.03 & 0.69 && 1.00  & 0.03 & 0.02 & 0.77 && 1.00  & 0.01 & 0.01 & 0.89 \\[0.2cm]
  $\alpha_{0}$  & 0.00  & -0.05 & 0.27 & 0.25 & 0.92 && -0.02 & 0.18 & 0.17 & 0.94 && -0.01 & 0.08 & 0.07 & 0.93 \\
  $\alpha_{1}$  & 0.50  & 0.51  & 0.10 & 0.09 & 0.94 && 0.50  & 0.07 & 0.06 & 0.94 && 0.50  & 0.03 & 0.03 & 0.94 \\ 
  $\alpha_{2}$  & 1.00  & 1.02  & 0.09 & 0.09 & 0.93 && 1.01  & 0.07 & 0.06 & 0.94 && 1.00  & 0.03 & 0.03 & 0.95 \\ 
  $\alpha_{3}$  & 0.50  & 0.51  & 0.11 & 0.10 & 0.93 && 0.51  & 0.08 & 0.07 & 0.94 && 0.50  & 0.03 & 0.03 & 0.95 \\ 
  $\alpha_{4}$  & 1.00  & 1.02  & 0.10 & 0.09 & 0.92 && 1.01  & 0.06 & 0.06 & 0.94 && 1.00  & 0.03 & 0.03 & 0.95 \\ 
  $\alpha_{7}$  & 0.50  & 0.51  & 0.10 & 0.09 & 0.95 && 0.51  & 0.06 & 0.06 & 0.95 && 0.50  & 0.03 & 0.03 & 0.95 \\ 
  $\alpha_{8}$  & 1.00  & 1.02  & 0.10 & 0.09 & 0.93 && 1.01  & 0.06 & 0.06 & 0.95 && 1.00  & 0.03 & 0.03 & 0.96 \\[0.2cm]
  $\nu_{0}$     & -0.22 & -0.22 & 0.11 & 0.10 & 0.90 && -0.22 & 0.07 & 0.07 & 0.92 && -0.22 & 0.03 & 0.03 & 0.92 \\
  \bottomrule
  \end{tabular}
\end{subtable}

\bigskip

\begin{subtable}{\textwidth}
    \centering
    \subcaption{}
    \vspace{-0.1cm}
\begin{tabular}{@{}c@{~~}c@{~~~} c@{~~}c@{~~}c@{~~}c@{~~} c@{~~~} c@{~~}c@{~~}c@{~~}c@{~~} c@{~~~} c@{~~}c@{~~}c@{~~}c@{}}
\toprule
\multicolumn{2}{l}{$\kappa=1.33$} & \multicolumn{4}{c}{$n = 500$} && \multicolumn{4}{c}{$n = 1000$} && \multicolumn{4}{c}{$n = 5000$} \\
\cmidrule(r){3-6} \cmidrule(r){8-11} \cmidrule(){13-16}
{} & $\theta$ & $\hat{\theta}$ & SE & SEE & CP && $\hat{\theta}$ & SE & SEE & CP && $\hat{\theta}$ & SE & SEE & CP \\
  \midrule
  $\beta_{0}$   & 0.00 &  -0.00 & 0.06 & 0.05 & 0.87  && -0.00 & 0.04 & 0.04 & 0.90  && 0.00  & 0.02 & 0.02 & 0.94 \\
  $\beta_{1}$   & 1.00 &  1.00  & 0.04 & 0.04 & 0.91  && 1.00  & 0.03 & 0.03 & 0.94  && 1.00  & 0.01 & 0.01 & 0.94 \\ 
  $\beta_{2}$   & 0.50 &  0.50  & 0.03 & 0.03 & 0.89  && 0.50  & 0.02 & 0.02 & 0.91  && 0.50  & 0.01 & 0.01 & 0.93 \\ 
  $\beta_{3}$   & 0.50 &  0.50  & 0.03 & 0.03 & 0.93  && 0.50  & 0.02 & 0.02 & 0.95  && 0.50  & 0.01 & 0.01 & 0.94 \\ 
  $\beta_{4}$   & 1.00 &  1.00  & 0.03 & 0.03 & 0.88  && 1.00  & 0.02 & 0.02 & 0.92  && 1.00  & 0.01 & 0.01 & 0.94 \\ 
  $\beta_{5}$   & 0.50 &  0.50  & 0.03 & 0.03 & 0.91  && 0.50  & 0.02 & 0.02 & 0.94  && 0.50  & 0.01 & 0.01 & 0.96 \\ 
  $\beta_{6}$   & 1.00 &  1.00  & 0.03 & 0.03 & 0.88  && 1.00  & 0.02 & 0.02 & 0.91  && 1.00  & 0.01 & 0.01 & 0.94 \\[0.2cm]
  $\alpha_{0}$  & 0.00 & -0.01 & 0.20 & 0.18 & 0.93  && -0.01 & 0.13 & 0.13 & 0.95  && -0.00 & 0.06 & 0.06 & 0.94 \\
  $\alpha_{1}$  & 0.50 & 0.51  & 0.08 & 0.08 & 0.94  && 0.50  & 0.06 & 0.05 & 0.94  && 0.50  & 0.02 & 0.02 & 0.94 \\ 
  $\alpha_{2}$  & 1.00 & 1.02  & 0.08 & 0.08 & 0.93  && 1.01  & 0.06 & 0.05 & 0.94  && 1.00  & 0.02 & 0.02 & 0.95 \\ 
  $\alpha_{3}$  & 0.50 & 0.51  & 0.09 & 0.09 & 0.94  && 0.51  & 0.07 & 0.06 & 0.94  && 0.50  & 0.03 & 0.03 & 0.95 \\ 
  $\alpha_{4}$  & 1.00 & 1.02  & 0.08 & 0.08 & 0.92  && 1.01  & 0.05 & 0.05 & 0.94  && 1.00  & 0.02 & 0.02 & 0.95 \\ 
  $\alpha_{7}$  & 0.50 & 0.50  & 0.08 & 0.08 & 0.95  && 0.50  & 0.05 & 0.05 & 0.95  && 0.50  & 0.02 & 0.02 & 0.95 \\ 
  $\alpha_{8}$  & 1.00 & 1.02  & 0.08 & 0.08 & 0.93  && 1.01  & 0.05 & 0.05 & 0.95  && 1.00  & 0.02 & 0.02 & 0.96 \\[0.2cm]
  $\nu_{0}$     & 0.13 & 0.16  & 0.11 & 0.10 & 0.92  && 0.14  & 0.07 & 0.07 & 0.92  && 0.13  & 0.03 & 0.03 & 0.93 \\
  \bottomrule
  \end{tabular}
\end{subtable}
\end{table}


\begin{table}[h!]
\ContinuedFloat
\begin{subtable}{\textwidth}
    \centering
    \subcaption{}
\begin{tabular}{@{}c@{~~}c@{~~~} c@{~~}c@{~~}c@{~~}c@{~~} c@{~~~} c@{~~}c@{~~}c@{~~}c@{~~} c@{~~~} c@{~~}c@{~~}c@{~~}c@{}}
\toprule
\multicolumn{2}{l}{$\kappa=1.67$} & \multicolumn{4}{c}{$n = 500$} && \multicolumn{4}{c}{$n = 1000$} && \multicolumn{4}{c}{$n = 5000$} \\
\cmidrule(r){3-6} \cmidrule(r){8-11} \cmidrule(){13-16}
{} & $\theta$ & $\hat{\theta}$ & SE & SEE & CP && $\hat{\theta}$ & SE & SEE & CP && $\hat{\theta}$ & SE & SEE & CP \\
  \midrule
  $\beta_{0}$   & 0.00 & -0.00 & 0.05 & 0.05 & 0.90 && -0.00 & 0.03 & 0.03 & 0.92 && 0.00  & 0.01 & 0.01 & 0.94 \\
  $\beta_{1}$   & 1.00 & 1.00  & 0.04 & 0.03 & 0.92 && 1.00  & 0.03 & 0.02 & 0.95 && 1.00  & 0.01 & 0.01 & 0.94 \\ 
  $\beta_{2}$   & 0.50 & 0.50  & 0.03 & 0.03 & 0.91 && 0.50  & 0.02 & 0.02 & 0.94 && 0.50  & 0.01 & 0.01 & 0.94 \\ 
  $\beta_{3}$   & 0.50 & 0.50  & 0.02 & 0.02 & 0.94 && 0.50  & 0.02 & 0.02 & 0.95 && 0.50  & 0.01 & 0.01 & 0.94 \\ 
  $\beta_{4}$   & 1.00 & 1.00  & 0.03 & 0.02 & 0.92 && 1.00  & 0.02 & 0.02 & 0.94 && 1.00  & 0.01 & 0.01 & 0.95 \\ 
  $\beta_{5}$   & 0.50 & 0.50  & 0.02 & 0.02 & 0.94 && 0.50  & 0.02 & 0.02 & 0.94 && 0.50  & 0.01 & 0.01 & 0.96 \\ 
  $\beta_{6}$   & 1.00 & 1.00  & 0.03 & 0.03 & 0.90 && 1.00  & 0.02 & 0.02 & 0.93 && 1.00  & 0.01 & 0.01 & 0.94 \\[0.2cm]
  $\alpha_{0}$  & 0.00 & -0.00 & 0.16 & 0.15 & 0.93 && -0.00 & 0.11 & 0.10 & 0.94 && -0.00 & 0.05 & 0.05 & 0.95 \\
  $\alpha_{1}$  & 0.50 & 0.50  & 0.07 & 0.07 & 0.94 && 0.50  & 0.05 & 0.05 & 0.95 && 0.50  & 0.02 & 0.02 & 0.94 \\ 
  $\alpha_{2}$  & 1.00 & 1.02  & 0.07 & 0.07 & 0.92 && 1.01  & 0.05 & 0.05 & 0.93 && 1.00  & 0.02 & 0.02 & 0.95 \\ 
  $\alpha_{3}$  & 0.50 & 0.51  & 0.08 & 0.08 & 0.94 && 0.50  & 0.06 & 0.05 & 0.94 && 0.50  & 0.02 & 0.02 & 0.95 \\ 
  $\alpha_{4}$  & 1.00 & 1.02  & 0.07 & 0.07 & 0.92 && 1.01  & 0.05 & 0.05 & 0.94 && 1.00  & 0.02 & 0.02 & 0.95 \\ 
  $\alpha_{7}$  & 0.50 & 0.50  & 0.07 & 0.07 & 0.94 && 0.50  & 0.05 & 0.05 & 0.95 && 0.50  & 0.02 & 0.02 & 0.95 \\ 
  $\alpha_{8}$  & 1.00 & 1.01  & 0.07 & 0.07 & 0.93 && 1.01  & 0.05 & 0.05 & 0.95 && 1.00  & 0.02 & 0.02 & 0.96 \\[0.2cm]
  $\nu_{0}$     & 0.38 & 0.43  & 0.12 & 0.11 & 0.92 && 0.40  & 0.08 & 0.07 & 0.94 && 0.38  & 0.03 & 0.03 & 0.95 \\
  \bottomrule
  \end{tabular}
\end{subtable}

\bigskip

\begin{subtable}{\textwidth}
    \centering
    \subcaption{}
\begin{tabular}{@{}c@{~~}c@{~~~} c@{~~}c@{~~}c@{~~}c@{~~} c@{~~~} c@{~~}c@{~~}c@{~~}c@{~~} c@{~~~} c@{~~}c@{~~}c@{~~}c@{}}
\toprule
\multicolumn{2}{l}{$\kappa=2$} & \multicolumn{4}{c}{$n = 500$} && \multicolumn{4}{c}{$n = 1000$} && \multicolumn{4}{c}{$n = 5000$} \\
\cmidrule(r){3-6} \cmidrule(r){8-11} \cmidrule(){13-16}
{} & $\theta$ & $\hat{\theta}$ & SE & SEE & CP && $\hat{\theta}$ & SE & SEE & CP && $\hat{\theta}$ & SE & SEE & CP \\
  \midrule
  $\beta_{0}$   & 0.00 & -0.00 & 0.05 & 0.04 & 0.90 && -0.00 & 0.03 & 0.03 & 0.93 && -0.00 & 0.01 & 0.01 & 0.95 \\
  $\beta_{1}$   & 1.00 & 1.00  & 0.03 & 0.03 & 0.92 && 1.00  & 0.02 & 0.02 & 0.95 && 1.00  & 0.01 & 0.01 & 0.95 \\ 
  $\beta_{2}$   & 0.50 & 0.50  & 0.03 & 0.03 & 0.90 && 0.50  & 0.02 & 0.02 & 0.92 && 0.50  & 0.01 & 0.01 & 0.93 \\ 
  $\beta_{3}$   & 0.50 & 0.50  & 0.02 & 0.02 & 0.93 && 0.50  & 0.01 & 0.01 & 0.95 && 0.50  & 0.01 & 0.01 & 0.94 \\ 
  $\beta_{4}$   & 1.00 & 1.00  & 0.02 & 0.02 & 0.91 && 1.00  & 0.02 & 0.02 & 0.93 && 1.00  & 0.01 & 0.01 & 0.95 \\ 
  $\beta_{5}$   & 0.50 & 0.50  & 0.02 & 0.02 & 0.95 && 0.50  & 0.01 & 0.01 & 0.94 && 0.50  & 0.01 & 0.01 & 0.96 \\ 
  $\beta_{6}$   & 1.00 & 1.00  & 0.03 & 0.02 & 0.89 && 1.00  & 0.02 & 0.02 & 0.92 && 1.00  & 0.01 & 0.01 & 0.94 \\[0.2cm]
  $\alpha_{0}$  & 0.00 & -0.00 & 0.14 & 0.13 & 0.93 && -0.00 & 0.09 & 0.09 & 0.94 && -0.00 & 0.04 & 0.04 & 0.94 \\
  $\alpha_{1}$  & 0.50 & 0.50  & 0.07 & 0.06 & 0.94 && 0.50  & 0.05 & 0.04 & 0.94 && 0.50  & 0.02 & 0.02 & 0.94 \\ 
  $\alpha_{2}$  & 1.00 & 1.02  & 0.07 & 0.06 & 0.92 && 1.01  & 0.05 & 0.04 & 0.93 && 1.00  & 0.02 & 0.02 & 0.95 \\ 
  $\alpha_{3}$  & 0.50 & 0.51  & 0.07 & 0.07 & 0.94 && 0.51  & 0.05 & 0.05 & 0.94 && 0.50  & 0.02 & 0.02 & 0.95 \\ 
  $\alpha_{4}$  & 1.00 & 1.02  & 0.07 & 0.06 & 0.93 && 1.01  & 0.04 & 0.04 & 0.94 && 1.00  & 0.02 & 0.02 & 0.95 \\ 
  $\alpha_{7}$  & 0.50 & 0.50  & 0.06 & 0.06 & 0.94 && 0.50  & 0.04 & 0.04 & 0.95 && 0.50  & 0.02 & 0.02 & 0.95 \\ 
  $\alpha_{8}$  & 1.00 & 1.01  & 0.07 & 0.06 & 0.92 && 1.01  & 0.04 & 0.04 & 0.94 && 1.00  & 0.02 & 0.02 & 0.96 \\[0.2cm]
  $\nu_{0}$     & 0.59 & 0.65  & 0.13 & 0.12 & 0.91 && 0.62  & 0.08 & 0.08 & 0.93 && 0.59  & 0.03 & 0.03 & 0.95 \\
  \bottomrule
  \multicolumn{16}{p{0.86\textwidth}}{\footnotesize SE, standard deviation of estimates over 1000 replications; SEE, average of estimated standard errors over 1000 replications; CP, the empirical coverage probability of a nominal 95\% confidence interval.}\\
  \end{tabular}
\end{subtable}
\end{table}

\clearpage

\subsection{Bootstrapping Coverage Probabilities} \label{app:sgnd_boots}
Table~\ref{tab:sgnd_boots} contains the estimation and inference metrics, similar to those presented in Table~3 of the main paper. The standard errors in Table~\ref{tab:sgnd_boots} are estimated by bootstrapping with 100 bootstrap resamples over 200 replicates. The results are for $\kappa = 1$, $\tau = 0.15$ and $n=500$. Bootstrapping is an alternative approach to estimate the standard errors when standard error breakdown becomes an issue.

\begin{table}[h]
\caption{\label{tab:sgnd_boots}Bootstrap results: estimation and inference metrics}
\centering
\begin{tabular}{@{}c@{~~}c@{~~~}c@{~~}c@{~~}c@{~~}c@{}}
\toprule
\multicolumn{6}{l}{$\kappa=1$}\\
{} & {} & \multicolumn{4}{c}{$n = 500$} \\
\cmidrule(l){3-6}
{} & $\theta$ & $\hat{\theta}$ & SE & SEE & CP \\
  \midrule
  $\beta_{0}$   & 0.00  &  -0.00 & 0.07 & 0.11 & 0.98 \\
  $\beta_{1}$   & 1.00  &  1.00  & 0.05 & 0.07 & 0.97 \\ 
  $\beta_{2}$   & 0.50  &  0.50  & 0.04 & 0.06 & 0.98 \\ 
  $\beta_{3}$   & 0.50  &  0.50  & 0.04 & 0.05 & 0.98 \\ 
  $\beta_{4}$   & 1.00  &  1.00  & 0.04 & 0.05 & 0.98 \\ 
  $\beta_{5}$   & 0.50  &  0.50  & 0.04 & 0.05 & 0.98 \\ 
  $\beta_{6}$   & 1.00  &  1.00  & 0.05 & 0.06 & 0.96 \\[0.2cm]
  $\alpha_{0}$  & 0.00  &  -0.06 & 0.29 & 0.71 & 0.91 \\
  $\alpha_{1}$  & 0.50  &  0.50  & 0.09 & 0.11 & 0.96 \\ 
  $\alpha_{2}$  & 1.00  &  1.02  & 0.10 & 0.12 & 0.92 \\ 
  $\alpha_{3}$  & 0.50  &  0.51  & 0.10 & 0.15 & 0.94 \\ 
  $\alpha_{4}$  & 1.00  &  1.02  & 0.09 & 0.11 & 0.88 \\ 
  $\alpha_{7}$  & 0.50  &  0.51  & 0.10 & 0.12 & 0.94 \\ 
  $\alpha_{8}$  & 1.00  &  1.01  & 0.09 & 0.11 & 0.94 \\[0.2cm]
  $\nu_{0}$     & -0.22 &  -0.22 & 0.12 & 0.23 & 0.95 \\
  \bottomrule
  \multicolumn{6}{p{0.35\textwidth}}{\footnotesize SE, standard deviation of estimates over 200 replications; SEE, average of estimated standard errors over 200 replications; CP, the empirical coverage probability of a nominal 95\% confidence interval.}\\
 \end{tabular}
 \end{table}

\clearpage
\section{Real Data Analyses: Additional Results} \label{app:sgnd_data}
This section contains estimation metrics for the additional methods performed in Section~4.
\begin{itemize}
    \item Table~\ref{tab:sgnd_dataset_estimates_hprice_other} is analogous to Table~5 of the main paper, but showing the estimation metrics for the additional methods used in analyzing the Boston housing data.
    \item Table~\ref{tab:sgnd_dataset_estimates_diabetes_other} is analogous to Table~7 of the main paper, but showing the estimation metrics for the additional methods used in analyzing the diabetes data.
\end{itemize}

\begin{table}[h!]
\caption{\label{tab:sgnd_dataset_estimates_hprice_other}Boston Housing Data: estimation metrics}
\begin{subtable}{\textwidth}
    \centering
    \subcaption{}
    \vspace{-0.1cm}
    \begin{tabular}{@{}l@{~~}  r@{~}c@{~~}   r@{~}c@{~~}   c@{~~}  r@{~~}r   c@{~~}    r@{~~}r@{}}
    \toprule
{} & \multicolumn{4}{c}{GAMLSS-STEP} && \multicolumn{2}{c}{\begin{tabular}{@{}c@{}}GAMLSS \\ -BOOST\end{tabular}} && \multicolumn{2}{c}{BAMLSS} \\

\cmidrule(r){2-5} \cmidrule(r){7-8} \cmidrule(){10-11}

$\hat \nu_0$ & \multicolumn{4}{c}{0.48 (0.10)} && \multicolumn{2}{c}{0.60} && \multicolumn{2}{c}{0.41} \\

{} & \multicolumn{2}{c}{$\hat\beta_j$} & \multicolumn{2}{c}{$\hat\alpha_j$} && \multicolumn{1}{c}{$\hat\beta_j$}& \multicolumn{1}{c}{$\hat\alpha_j$} && \multicolumn{1}{c}{$\hat\beta_j$} & \multicolumn{1}{c}{$\hat\alpha_j$} \\
 \midrule
      \texttt{intercept}     &   3.76  & (0.08) & -7.96  & (1.96)  && 4.88  & -3.71  && 3.50    & -5.56   \\ 
      \texttt{ltax}      &   -0.19 & (0.01) & 0.84   & (0.33)  && -0.20 & 0.21   && -0.17   & 0.74    \\ 
      \texttt{rm}        &   0.21  & (0.01) &        &         && 0.11  &        && 0.24    & -0.30   \\ 
      \texttt{ldis}      &   -0.16 & (0.02) & -0.83  & (0.17)  &&       & -0.62  && -0.15   & -1.05   \\ 
      \texttt{llstat}    &   -0.21 & (0.02) &        &         && -0.35 &        && -0.17   & 0.01    \\ 
      \texttt{ptratio}   &   -0.03 & (0.00) &        &         && -0.02 &        && -0.02   & 0.00    \\ 
      \texttt{crim}      &   -0.02 & (0.00) &        &         &&       & 0.03   && -0.02   & 0.02    \\ 
      \texttt{rad}       &   0.01  & (0.00) & 0.05   & (0.02)  &&       & 0.03   && 0.01    & 0.05    \\ 
      \texttt{lnox}      &   -0.39 & (0.07) &        &         &&       &        && -0.24   & -0.41   \\ 
      \texttt{age}       &         &        &        &         &&       &        && -0.00   & 0.00    \\ 
      \texttt{chast}     &   0.05  & (0.02) &        &         &&       & 0.04   && 0.04    & 0.25    \\ 
      \texttt{zn}        &         &        &        &         && -0.00 & 0.01   && -0.00   & 0.00    \\ 
      \texttt{indus}     &         &        &        &         &&       & -0.01  && -0.00   & -0.03   \\ 

\bottomrule
\end{tabular}
\end{subtable}

\bigskip

\begin{subtable}{\textwidth}
    \centering
    \subcaption{}
    \vspace{-0.1cm}
\begin{tabular}{@{}l@{~~}  r@{~}c@{~~}   r@{~}c@{~~}   c@{~~}  r@{~~}r   c@{~~}    r@{~~}r@{}}
    \toprule
    
{} & \multicolumn{4}{c}{SGND-SPR} && \multicolumn{2}{c}{ALASSO} && \multicolumn{2}{c}{{\begin{tabular}{@{}c@{}}LAD \\ -LASSO\end{tabular}}} \\

\cmidrule(r){2-5} \cmidrule(r){7-8} \cmidrule(){10-11}

$\hat \nu_0$ & \multicolumn{4}{c}{-0.42 (0.30)} && \multicolumn{2}{c}{} && \multicolumn{2}{c}{} \\

{} & \multicolumn{2}{c}{$\hat\beta_j$} & \multicolumn{2}{c}{$\hat\alpha_j$} && \multicolumn{1}{c}{$\hat\beta_j$}& \multicolumn{1}{c}{$\hat\alpha_j$} && \multicolumn{1}{c}{$\hat\beta_j$} & \multicolumn{1}{c}{$\hat\alpha_j$} \\
 \midrule
      \texttt{intercept}     &  3.85  & (0.56) & -4.59 & (0.94) && 5.27  & -3.39 && 3.94  & -4.13  \\ 
      \texttt{ltax}      &  -0.14 & (0.06) &       &        && -0.21 &       && -0.13 &        \\ 
      \texttt{rm}        &  0.17  & (0.05) &       &        && 0.06  &       && 0.16  &        \\ 
      \texttt{ldis}      &  -0.19 & (0.05) &       &        && -0.22 &       && -0.16 &        \\ 
      \texttt{llstat}    &  -0.26 & (0.06) &       &        && -0.40 &       && -0.27 &        \\ 
      \texttt{ptratio}   &  -0.03 & (0.01) &       &        && -0.03 &       && -0.03 &        \\ 
      \texttt{crim}      &  -0.01 & (0.01) &       &        && -0.01 &       && -0.01 &        \\ 
      \texttt{rad}       &  0.01  & (0.01) &       &        && 0.01  &       && 0.00  &        \\ 
      \texttt{lnox}      &  -0.46 & (0.17) &       &        && -0.45 &       && -0.37 &        \\ 
      \texttt{age}       &        &        &       &        &&       &       &&       &        \\ 
      \texttt{chast}     &  0.06  & (0.05) &       &        && 0.08  &       && 0.08  &        \\ 
      \texttt{zn}        &        &        &       &        &&       &       && -0.00 &        \\ 
      \texttt{indus}     &        &        &       &        &&       &       &&       &        \\
\bottomrule
\multicolumn{11}{p{0.7\textwidth}}{\footnotesize SGND-SPR standard errors estimated by bootstrapping with 200 bootstrap resamples.}
\end{tabular}
\end{subtable}
\end{table}

\begin{table}[h!]
\caption{\label{tab:sgnd_dataset_estimates_diabetes_other}Diabetes Data: estimation metrics}
\begin{subtable}{\textwidth}
    \centering
    \subcaption{}
    \vspace{-0.1cm}
    \begin{tabular}{@{}l@{~~}  r@{~}c@{~~}   r@{~}c@{~~}   c@{~~}  r@{~~}r   c@{~~}    r@{~~}r@{}}
    \toprule
{} & \multicolumn{4}{c}{GAMLSS-STEP} && \multicolumn{2}{c}{\begin{tabular}{@{}c@{}}GAMLSS \\ -BOOST\end{tabular}} && \multicolumn{2}{c}{BAMLSS} \\

\cmidrule(r){2-5} \cmidrule(r){7-8} \cmidrule(){10-11}

$\hat \nu_0$ & \multicolumn{4}{c}{0.97 (0.14)} && \multicolumn{2}{c}{1.56} && \multicolumn{2}{c}{0.82} \\

{} & \multicolumn{2}{c}{$\hat\beta_j$} & \multicolumn{2}{c}{$\hat\alpha_j$} && \multicolumn{1}{c}{$\hat\beta_j$}& \multicolumn{1}{c}{$\hat\alpha_j$} && \multicolumn{1}{c}{$\hat\beta_j$} & \multicolumn{1}{c}{$\hat\alpha_j$} \\
 \midrule
        \texttt{intercept} & -185.11 & (35.46) & 7.30 & (0.45) && 81.52  & 7.59  && -27.99  & 12.81  \\ 
        \texttt{BMI}   & 5.36    & (0.72)  & 0.06 & (0.02) && 1.02   & 0.05  && 4.75    & 0.02   \\
        \texttt{S5}    & 40.98   & (5.72)  &      &        && 9.45   &       && 0.20    & -0.82  \\
        \texttt{S3}    & -1.04   & (0.23)  &      &        &&        &       && -3.10   & -0.08  \\
        \texttt{BP}    & 0.93    & (0.22)  &      &        &&        & 0.00  && 1.08    & 0.02   \\
        \texttt{SEX}   & -20.70  & (5.88)  &      &        &&        &       && -25.70  & -0.49  \\
        \texttt{S1}    &         &         &      &        &&        &       && 2.02    & 0.04   \\
        \texttt{S2}    &         &         &      &        &&        & -0.00 && -2.27   & -0.04  \\
        \texttt{S4}    &         &         &      &        &&        &       && 1.26    & -0.28  \\
        \texttt{S6}    &         &         &      &        &&        & 0.01  && 0.14    & 0.00   \\ 
        \texttt{AGE}   &         &         &      &        &&        &       && 0.16    & -0.01  \\
\bottomrule
\end{tabular}
\end{subtable}

\bigskip

\begin{subtable}{\textwidth}
    \centering
    \subcaption{}
    \vspace{-0.1cm}
\begin{tabular}{@{}l@{~~}  r@{~}c@{~~}   r@{~}c@{~~}   c@{~~}  r@{~~}r   c@{~~}    r@{~~}r@{}}
    \toprule
    
{} & \multicolumn{4}{c}{SGND-SPR} && \multicolumn{2}{c}{ALASSO} && \multicolumn{2}{c}{{\begin{tabular}{@{}c@{}}LAD \\ -LASSO\end{tabular}}} \\

\cmidrule(r){2-5} \cmidrule(r){7-8} \cmidrule(){10-11}

$\hat \nu_0$ & \multicolumn{4}{c}{0.66 (0.13)} && \multicolumn{2}{c}{} && \multicolumn{2}{c}{} \\

{} & \multicolumn{2}{c}{$\hat\beta_j$} & \multicolumn{2}{c}{$\hat\alpha_j$} && \multicolumn{1}{c}{$\hat\beta_j$}& \multicolumn{1}{c}{$\hat\alpha_j$} && \multicolumn{1}{c}{$\hat\beta_j$} & \multicolumn{1}{c}{$\hat\alpha_j$} \\
 \midrule
        \texttt{intercept} &  -310.15 & (26.17) & 8.54  & (0.12) && -296.86 & 7.98  && -247.22 & 8.01  \\ 
        \texttt{BMI}   &  5.76    & (0.69)  &       &        && 5.83    &       && 4.83    &       \\
        \texttt{S5}    &  71.46   & (7.73)  &       &        && 61.88   &       && 52.17   &       \\
        \texttt{S3}    &          &         &       &        &&         &       && -0.82   &       \\
        \texttt{BP}    &  1.09    & (0.21)  &       &        && 1.08    &       && 1.23    &       \\
        \texttt{SEX}   &  -19.49  & (5.92)  &       &        && -19.21  &       && -27.84  &       \\
        \texttt{S1}    &  -1.02   & (0.23)  &       &        && -0.70   &       && -0.12   &       \\
        \texttt{S2}    &  0.85    & (0.24)  &       &        && 0.41    &       &&         &       \\
        \texttt{S4}    &          &         &       &        && 4.55    &       &&         &       \\
        \texttt{S6}    &          &         &       &        &&         &       && 0.20    &       \\
        \texttt{AGE}   &          &         &       &        &&         &       &&         &       \\ 
\bottomrule
\end{tabular}
\end{subtable}
\end{table}

\end{document}